\documentclass[preprint,showpacs,preprintnumbers,amsmath,amssymb]{revtex4}
\usepackage{url} 
\newcommand{\eqnref}[1]{equation \eqref{#1}}
\newcommand{\eqnrefs}[1]{equations \eqref{#1}}
\newcommand{\figref}[1]{Figure \ref{#1}}
\newcommand{\figrefs}[1]{Figures \ref{#1}}
\newcommand{\Eqnref}[1]{Equation \eqref{#1}}

\newcommand{\Figref}[1]{Figure \ref{#1}}
\newcommand{\Figrefs}[1]{Figures \ref{#1}}
\usepackage{amsmath}
\usepackage{amsfonts}
\usepackage{tikz}
\usepackage{graphicx}
\usepackage{epstopdf}
\usepackage{epsf}
\usepackage{setspace}

\usepackage[paperwidth=210mm,paperheight=297mm,centering,hmargin=1.8cm,vmargin=2.3cm]{geometry}

\graphicspath{{pdf/}{Figures/}}

\begin{document}

\title{Virtual sources and receivers in the real Earth - considerations for practical applications}

\author{Joeri Brackenhoff, Jan Thorbecke and Kees Wapenaar \\
\textit{Department of Geoscience and Engineering, Delft University of Technology,} \\
\textit{P.O. Box 5048, 2600 GA Delft, The Netherlands}}

\begin{abstract}
To enhance monitoring of the subsurface, virtual sources and receivers inside the subsurface can be created from seismic reflection data at the surface of the Earth using the Marchenko method. The response between these virtual sources and receivers can be obtained through the use of homogeneous Green's function retrieval. A homogeneous Green's function is a superposition of a Green's function and its time-reversal. The main aim of this paper is to obtain accurate homogeneous Green's functions from field data. Classical homogeneous Green's function retrieval requires an unrealistic enclosing recording surface, however, by using a recently proposed single-sided retrieval scheme, this requirement can be avoided. We first demonstrate the principles of using the single-sided representation on synthetic data and show that different source signatures can be taken into account. Because the Marchenko method is sensitive to recording limitations of the reflection data, we study five cases of recording limitations with synthetic data and demonstrate their effects on the final result. Finally, the method is demonstrated on a pre-processed field dataset which fulfills the requirements for applying the single-sided Green's function retrieval scheme. The scheme has the potential to be used in future applications, such as source localization.
\end{abstract}

\maketitle

\section{Introduction}
Seismic data can be used in a variety of ways to monitor and explore the subsurface of the Earth. Such data are obtained by measuring the wavefield that is propagating through the subsurface at physical receivers. Seismic data can be acquired using an active source at the surface of the Earth, in which case receivers are usually located on the same surface as the source, or in a borehole. The receivers measure the full wavefield, i.e., both primary and multiply scattered events. These measurements are often used to obtain information about the structure of the subsurface and its properties \citep{yilmaz2001seismic}. Alternatively, data can be acquired using a passive source, which is a source of the wavefield that occurs naturally in the subsurface of the Earth. In this setup, the wavefield is recorded by a continuously recording receiver array, usually at the surface of the Earth. These measurements can contain additional information about processes in the subsurface, such as induced seismicity \citep{grigoli2017current}. These types of measurements are receiving more attention because of the potentially damaging effects of induced seismicity in residential areas \citep{van2015induced,magnani2017discriminating}. \\
Active measurements can be employed to supplement the passive measurments. Using advanced seismic processing techniques, the wavefield that is measured at the surface of the Earth can be redatumed to locations inside the subsurface. By redatuming receivers from their physical location on the surface to locations at depth, virtual receivers are created. The advantage of such virtual receivers is that, by considering many of them, the evolution of the wavefield through the subsurface over time can be studied, which can provide relevant information about source mechanisms and the locations of scatterers in the subsurface. Similar to receiver redatuming, physical sources at the surface can be redatumed to create virtual sources at any location in the subsurface. Furthermore, the response between any combination of a virtual source and virtual receiver can be retrieved, a process we call homogeneous Green's function retrieval. Whereas a Green's function describes the response of a medium to a Dirac function, a homogeneous Green's function is a Green's function superposed by its time-reversal to avoid a source singularity. The classical representation for the homogeneous Green's function retrieval was derived by \citet{porter1970diffraction}. This method was further extended for inverse source problems by \citet{porter1982holography} and inverse scattering methods by \citet{oristaglio1989inverse}. This classical representation has been employed as the theoretical basis in the field of seismic interferometry to create virtual sources \citep{wapenaar2004retrieving,van2005modeling,bakulin2005virtual} or virtual receivers \citep{curtis2009virtual}. However, in all of these applications, it appeared that a complete enclosing boundary is vital for retrieving a full homogeneous Green's function without artifacts. \\
Recently, a new single-sided representation for homogeneous Green's function retrieval has been derived. Instead of an enclosing boundary, it uses a single, non-enclosing boundary, typically the Earth's surface \citep{wapenaar2017virtual}. An example of the application of this method on synthetic data can be found in \citet{wapenaar2016singleGH}. In this approach, the data-driven Marchenko method is used to create virtual sources and receivers in the subsurface from reflection data at the Earth's surface. Using the homogeneous Green's function retrieval, the response between one selected virtual source and all virtual receivers is obtained. The Marchenko method, for the purpose of geophysial applications, was first proposed for 1D by \cite{broggini2012focusing}, based on work by \cite{rose2001single}, and was later extended for 2D and 3D applications \citep{wapenaar2013three,wapenaar2014marchenko}. The method uses two types of input. The first is active-source single-sided seismic reflection data measured at the surface of the Earth. The second is an estimation of the first wavefield event, which is called the first arrival, that would be caused by a source from a location in the subsurface to receiver locations at the surface of the Earth (hence, the first arrival of a Green's function between a subsurface location and the surface). The locations of the receivers of the Green's function match the locations of the receivers of the reflection response. The Marchenko method uses these data to create a full waveform Green's function, including all multiple scattering, for a virtual source in the subsurface and receivers at the Earth's surface. To model the first arrival, only a background velocity model is required, which can be estimated by processing the reflection data. A dense array of virtual sources for Green's functions in the subsurface can be created through repeated use of this methodology. Aside from the Green's function, the Marchenko method is also capable of retrieving a focusing function, which is designed to focus from the single-sided surface, where the reflection response is measured, to a focal location in the subsurface without any reverberation artifacts. The single-sided representation uses the focusing function, together with a Green's function, to create the response between a virtual source and receiver. Due to the single-sided focusing properties of the focusing function, the retrieval can be done for a single-sided recording setup without any artifacts.\\
Employing the Marchenko method on field data for practical applications is challenging due to the sensitivity of the Marchenko method to recording limitations of the reflection response. The sensitivity is partially caused by the fact that in the derivation of the Marchenko method, evanescent waves are ignored and it is assumed that the medium of interest is lossless. In real media, the wavefield suffers from absorption, which violates the latter assumption. Furthermore, the method requires the reflection response to be well sampled and the aperture to be sufficiently large. The Marchenko method has been succesfully applied on field data, by pre-processing the reflection response. Examples for the purpose of imaging can be found in \citet{ravasi2016target} and \citet{staring2017adaptive}, who used adaptive corrections in the Marchenko method. Homogeneous Green's function retrieval using the single-sided representation on field data was achieved by \citet{wapenaar2018virtual} and \citet{brackenhoff2019}.\\
The aim of this paper is to apply the single-sided representation on field data and to consider the influence of recording limitations of the reflection response on the retrieved homogeneous Green's functions. To this end, we consider a 2D field seismic dataset from the V\o ring basin off the coast of Norway. Along with the field data, we also consider a subsurface model, that is designed to simulate the subsurface of the area where the actual reflection response is recorded. Using this model, synthetic reflection data are created. First, we use the synthetic data to make a comparison between the results that are obtained when the single-sided representation is used and when the classical representation is used. The results show that the homogeneous Green's function is more accurately retrieved when the single-sided representation is used. The first arrivals that are used in these tests in the Marchenko method are all modeled using a monopole source mechanism. To study the influence of the source mechanism on the final result, the experiment is repeated using first arrivals that were modeled using a double-couple source mechanism, which is more representative for small-scale earthquakes \citep{aki2002quantitative}. The homogeneous Green function that is obtained in this way still contains the correct events and has a double-couple signature. Next, we determine the sensitivity of the result to five recording limitations on the reflection data, namely coarse source-receiver sampling, missing near offsets, small aperture, offsets missing in one direction and absorption of the reflection data. The results of these numerical experiments are taken into account so that the field reflection data can be pre-processed and the single-sided representation can be applied properly. We employ both the classical and single-sided representation to the field data in order to compare the results. The applications show the potential of the single-sided representation for field data, as well as the possibility of applying the representation to passive field recordings.

\newpage

\section{Theory}
In this section, we present an overview of the definitions and equations that are required for homogeneous Green's function retrieval. The Green's function and focusing function are reviewed, followed by the definitions of the classical enclosed boundary and single-sided representations for homogeneous Green's function retrieval. The Marchenko method and its limitations are considered, as well as the double-couple source mechanism.
\begin{figure}[!hptb]
	\centering
		\includegraphics[width=\columnwidth]{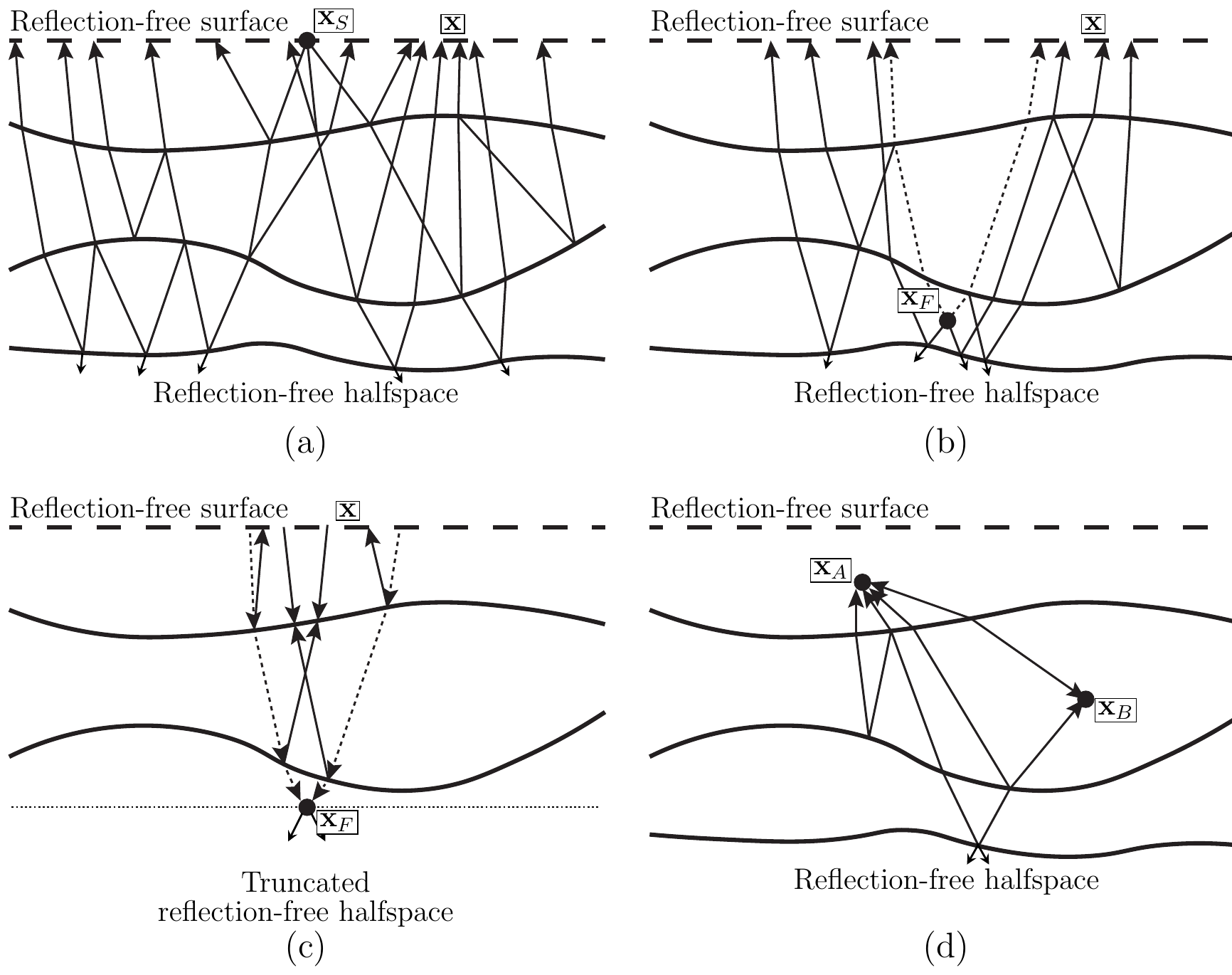}
	\caption{\textit{Possible raypaths drawn for, }(a)\textit{ a reflection response }$R(\textbf{x},\textbf{x}_S,t)$\textit{, measured at varying receiver locations }$\textbf{x}$\textit{ at the surface, with a source at }$\textbf{x}_S$\textit{ also at the surface, }(b)\textit{ a Green's function} $G(\textbf{x},\textbf{x}_F,t)$\textit{, measured at varying receiver location }$\textbf{x}$\textit{ at the surface with a source at }$\textbf{x}_F$\textit{ inside the medium, }(c)\textit{ a focusing function} $f_1(\mathbf{x},\mathbf{x}_F,t)$\textit{, emitted from the surface at varying locations }$\textbf{x}$\textit{, focusing to a focal location} $\textbf{x}_F$ \textit{inside a medium that is truncated below} $\textbf{x}_F$\textit{ (indicated by the horizontal dotted line), and }(d)\textit{ a homogeneous Green's function} $G_{\rm h}(\textbf{x}_A,\textbf{x}_B,t)$\textit{, between two locations, }$\textbf{x}_A$\textit{ and }$\textbf{x}_B$\textit{ inside the medium. The homogeneous Green's function is indicated with two-sided arrows to represent that it is the superposition of Green's function and its time-reversal. The dotted arrows in (b) and (c) indicate the first arrival for the Green's function and focusing function. The surfaces at the top of all figures are transparent, hence, there are no free-surface multiples.}}
	\label{functions}
\end{figure}
\subsection{Green's function}
The Green's function is defined as the solution of the wave equation to a Dirac point source which can be written as \citep{feynman2011feynman, Morse53Book}:
\begin{equation} \label{T1}
\begin{split}
\Bigg(\rho(\mathbf{x})\partial_i\Bigg(\frac{1}{\rho(\mathbf{x})}\partial_i\Bigg)-\frac{1}{c^2(\mathbf{x})}\partial^2_t\Bigg)G(\mathbf{x},\mathbf{x}_F,t)= -\rho(\mathbf{x})\delta(\mathbf{x}-\mathbf{x}_F)\partial_t\delta(t),
\end{split}
\end{equation}
where $G(\mathbf{x},\mathbf{x}_F,t)$ describes the response of the medium, at time $t$, at location $\mathbf{x}$ to a source at location $\mathbf{x}_F$. The locations are defined in 3D such that $\mathbf{x}=(x_1,x_2,x_3)^T$. The symbols $\rho$ and $c$ indicate the density and velocity of the medium, respectively, $\delta$ indicates a Dirac delta function, $\partial_t$ a temporal derivative and $\partial_i$ the partial derivative in the three principal directions. The repeated subscript $i$ follows the Einstein summation convention. Note that the source at the right hand side is defined with a temporal derivative acting on the Dirac delta function. This choice is made to simulate a volume injection-rate source. According to the reciprocity principle, the source and receiver location of the Green's functions can be interchanged, $G(\mathbf{x},\mathbf{x}_F,t)=G(\mathbf{x}_F,\mathbf{x},t)$.\\ 
We also consider the Fourier-transformed Green's function $G(\mathbf{x},\mathbf{x}_F,\omega)$:
\begin{equation} \label{TA2}
G(\mathbf{x},\mathbf{x}_F,\omega)=\int_{-\infty}^{\infty}G(\mathbf{x},\mathbf{x}_F,t)e^{i\omega t}\text{d}t,
\end{equation}
where $\omega$ denotes the angular frequency and $i$ the imaginary unit. Note, that the sign in the exponential can be reversed, as long as the same is done for the inverse Fourier transform. 
Using \eqnref{TA2}, \eqnref{T1} is transformed to the frequency domain:
\begin{equation} \label{TA1}
\begin{split}
\Bigg(\rho(\mathbf{x})\partial_i\Bigg(\frac{1}{\rho(\mathbf{x})}\partial_i\Bigg)+\frac{\omega^2}{c^2(\mathbf{x})}\Bigg)G(\mathbf{x},\mathbf{x}_F,\omega)= i\omega\rho(\mathbf{x})\delta(\mathbf{x}-\mathbf{x}_F).
\end{split}
\end{equation} 
A schematic illustration of the Green's function is shown in \figref{functions}(b), where the receivers are placed at the surface of a medium and its source inside the medium. We assume that the medium has a transparent reflection-free halfspace at the top of the medium. In practice this situtation is obtained after the elimination of surface-related multiples. Some possible raypaths, including scattering, have been drawn in the figure. \mbox{\Figref{functions}(a)} shows a special case of the Green's function, with both source and receivers placed at the surface of the medium. This is called the reflection response $R(\textbf{x},\textbf{x}_S,t)$ and contains all the reflections, both primaries and multiples, of the medium, however, we assume that the direct wave from the sources to the receivers are not present in the reflection response. \\
The homogeneous Green's function is defined as the superposition of the Green's function and its time-reversal. Because of the temporal derivative on the Dirac delta function in \eqnref{T1}, when time-reversal is applied, the source term will obtain an opposite sign. As a result, the superposition of the Green's function and its time-reversal removes the source term, thereby avoiding a singularity at the source position:
\begin{equation} \label{T21}
G_{\rm h}(\mathbf{x},\mathbf{x}_F,t)=G(\mathbf{x},\mathbf{x}_F,t)+G(\mathbf{x},\mathbf{x}_F,-t),
\end{equation}
\begin{equation} \label{T3}
\begin{split}
\Bigg(\rho(\mathbf{x})\partial_i\Bigg(\frac{1}{\rho(\mathbf{x})}\partial_i\Bigg)-\frac{1}{c^2(\mathbf{x})}\partial^2_t\Bigg)G_{\rm h}(\mathbf{x},\mathbf{x}_F,t)=0,
\end{split}
\end{equation}
and in the frequency domain:
\begin{equation} \label{TF21}
G_{\rm h}(\mathbf{x},\mathbf{x}_F,\omega)=G(\mathbf{x},\mathbf{x}_F,\omega)+G^*(\mathbf{x},\mathbf{x}_F,\omega),
\end{equation}
\begin{equation} \label{TF3}
\begin{split}
\Bigg(\rho(\mathbf{x})\partial_i\Bigg(\frac{1}{\rho(\mathbf{x})}\partial_i\Bigg)+\frac{\omega^2}{c^2(\mathbf{x})}\Bigg)G_{\rm h}(\mathbf{x},\mathbf{x}_F,\omega)=0,
\end{split}
\end{equation}
where $G_{\rm h}(\mathbf{x},\mathbf{x}_F,t)$ and $G_{\rm h}(\mathbf{x},\mathbf{x}_F,\omega)$ denote the homogeneous Green's function in the time domain and frequency domain, respectively, and the asterisk indicates complex conjugation. \Figref{functions}(d) shows a schematic illustration of the homogeneous Green's function, $G_{\rm h}(\mathbf{x}_A,\mathbf{x}_B,t)$, with some possible raypaths drawn, with both its source and receiver inside the medium. To reflect the superposition of the Green's function and its time-reversal, the raypaths are indicated with two-sided arrows.
\subsection{Focusing Function}
The focusing function $f_1(\mathbf{x},\mathbf{x}_F,t)$ describes a wavefield, at time $t$, at location $\mathbf{x}$, that focuses to a focal location $\mathbf{x}_F$ in the subsurface. The focusing function propagates in a medium that is truncated below $\mathbf{x}_F$, which means that there are no reflectors present below the focal location. \\
The focusing function can be decomposed into its upgoing and downgoing parts:
\begin{equation} \label{T4a}
f_1(\mathbf{x},\mathbf{x}_F,t)=f_1^+(\mathbf{x},\mathbf{x}_F,t)+f_1^{-}(\mathbf{x},\mathbf{x}_F,t),
\end{equation}
where $f_1^+(\mathbf{x},\mathbf{x}_F,\omega)$ denotes the downgoing focusing function and $f_1^-(\mathbf{x},\mathbf{x}_F,\omega)$ the upgoing focusing function. The downgoing part of the focusing function is defined as the inverse of the transmission response of the truncated medium \citep{wapenaar2014marchenko}.\\
The focusing function is schematically illustrated in \figref{functions}(c), where some possible raypaths have been drawn. The first arrival, which is indicated by the dotted raypath, propagates from the surface to the focal location and scatters at the reflectors, creating an upgoing wavefield. In order to ensure that these upgoing waves do not cause additional events arriving after the wavefield has focused, additional downgoing waves are injected from the surface, which cancel out these events. This occurs at the locations where arrowheads meet in \figref{functions}(c). Because of the reflection-free surface at $\mathbf{x}$, there are no events present in the focusing function to account for free-surface multiples. A more detailed description of the focusing function can be found in \citet{slob2014seismic}.
\subsection{Homogeneous Green's function representation}
\begin{figure}[!hptb]
	\centering
	\includegraphics[width=\columnwidth]{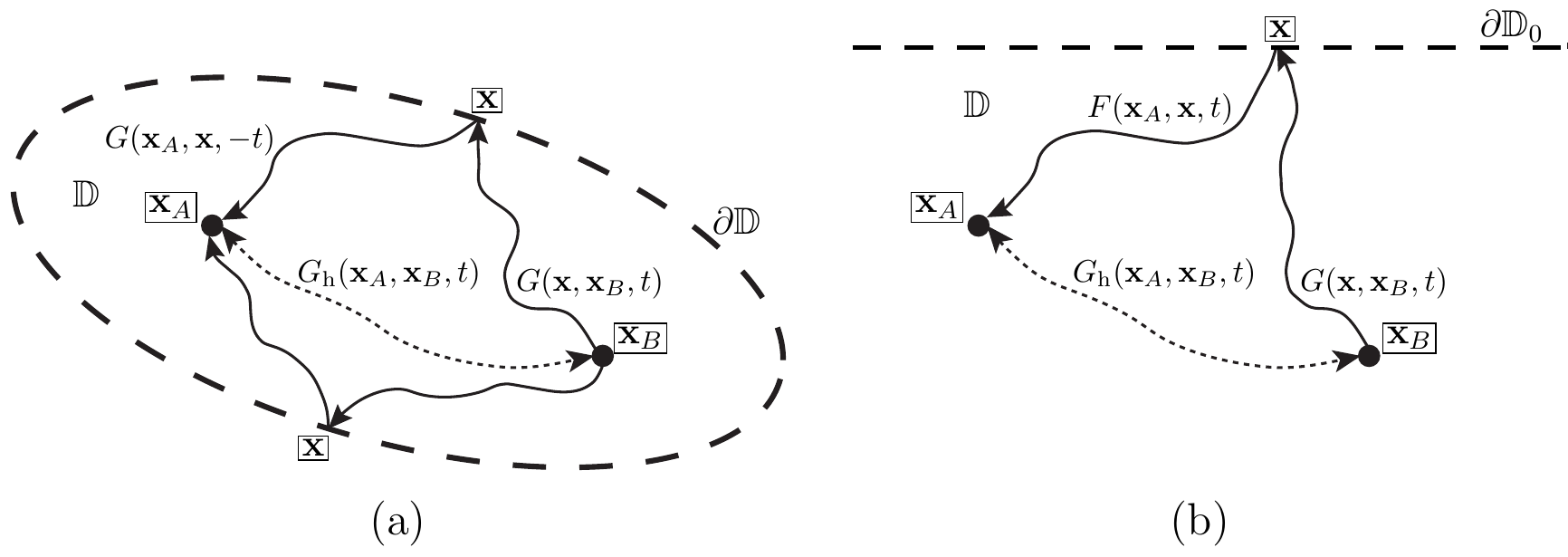}
	\caption{\textit{Recording setup for homogeneous Green's function retrieval using }(a)\textit{ the classical representation and }(b)\textit{ the single-sided representation. For both setups, a Green's function, }$G(\mathbf{x},\mathbf{x}_B,t)$\textit{ is utilized. For the classical representation, an additional Green's function,} $G(\mathbf{x}_A,\mathbf{x},t)$\textit{ is used to create a virtual receiver location. The medium of interest $\mathbb{D}$ is surrounded by an enclosing boundary $\partial\mathbb{D}$. \Eqnref{T5} is evaluated over this enclosing boundary. For the single-sided representation, the virtual receiver is not created using a Green's function, but rather a focusing function,} $F(\mathbf{x}_A,\mathbf{x},t)$\textit{. The medium $\mathbb{D}$ is not enclosed, instead only a single-sided non-enclosing boundary $\partial\mathbb{D}_0$ is present at a single side. \Eqnref{T6} is evaluated over this single-sided bounadry. The homogeneous Green's function is indicated by the dotted line.}}
	\label{representation}
\end{figure}
The classical representation of the homogeneous Green's function states that the response between a source and receiver inside a medium can be retrieved from observations at a boundary. In order to achieve this, an enclosing boundary around the medium of interest, over which the data can be recorded and/or injected, needs to be present \citep{porter1970diffraction,porter1982holography,oristaglio1989inverse}. The classical representation in the frequency domain can be written as follows:
\begin{equation} \label{T5}
\begin{split}
G_{\rm h}(\mathbf{x}_A,\mathbf{x}_B,\omega) =  \oint_{\partial\mathbb{D}}\frac{-1}{i\omega\rho(\mathbf{x})}\{&\partial_i G^*(\mathbf{x}_A,\mathbf{x},\omega)G(\mathbf{x},\mathbf{x}_B,\omega)\\
 -  &G^*(\mathbf{x}_A,\mathbf{x},\omega)\partial_iG(\mathbf{x},\mathbf{x}_B,\omega)\}n_i\text{d}^2\mathbf{x},
\end{split}
\end{equation}
where $n_i$ indicates the components of the normal vector in the three principal directions. The integral is evaluated over a boundary $\partial\mathbb{D}$ enclosing the medium $\mathbb{D}$. In \eqnref{T5}, the function $G(\mathbf{x},\mathbf{x}_B,\omega)$ describes the response of the medium at varying location $\textbf{x}$ at the boundary to a source at location $\textbf{x}_B$ inside the medium. The time-reversed function $G^*(\mathbf{x}_A,\mathbf{x},\omega)$ back-propagates the responses from the boundary to the receiver location $\mathbf{x}_A$, thereby creating a virtual receiver at $\mathbf{x}_A$. A schematic overview of the application of this representation is shown in \figref{representation}(a).\\
In practice, the classical representation is often not evaluated correctly, because acquisition on an enclosing boundary is not feasible and only measurements on a single-sided non-enclosing boundary, usually the Earth's surface, are available. As an approximation, \eqnref{T5} can be evaluated over the single-sided boundary. Applying the representation in this way causes significant artifacts in the retrieved homogeneous Green's function, however. Due to the fact that few alternatives are available, the method is still widely applied to cases where no closed boundaries are present.\\
An alternative representation that can be employed uses a focusing function instead of a Green's function. This representation is capable of retrieving the full homogeneous Green's function with significantly less artifacts from a single-sided boundary, hence it is referred to as the single-sided representation. It can be written as \cite[equation 30]{wapenaar2017virtual}:
\begin{equation} \label{T6}
G_{\rm h}(\mathbf{x}_A,\mathbf{x}_B,\omega) =  \mathfrak{I}\int_{\partial\mathbb{D}_0}\frac{4}{\omega\rho(\mathbf{x})}\{ \partial_3F(\mathbf{x}_A,\mathbf{x},\omega)G(\mathbf{x},\mathbf{x}_B,\omega)\}\text{d}^2\mathbf{x},
\end{equation}
where $\partial\mathbb{D}_0$ denotes the single-sided boundary and $\mathfrak{I}$ the imaginary part of a complex function. The focusing function, $F(\mathbf{x}_A,\mathbf{x},\omega)$, is defined as:
\begin{equation} \label{T4}
F(\mathbf{x}_A,\mathbf{x},\omega)=f_1^+(\mathbf{x},\mathbf{x}_A,\omega)-f_1^{-*}(\mathbf{x},\mathbf{x}_A,\omega).
\end{equation}
In \eqnref{T6}, $G(\mathbf{x},\mathbf{x}_B,\omega)$ serves again as the response to a source location inside the medium, measured at the single-sided boundary $\partial\mathbb{D}_0$. The focusing function $F(\mathbf{x}_A,\mathbf{x},\omega)$ serves as the back-propagator of the responses from the boundary to the focal location inside the medium. A schematic representation of this procedure is shown in \figref{representation}(b).\\
The two representations in \eqnrefs{T5} and \eqref{T6} for homogeneous Green's function retrieval are similar in form, as both use a backward propagator on the response measured on the boundary. The main difference is that, for the single-sided representation, the backward propagator is a focusing function instead of a time-reversed Green's function. As one can interpret from \figref{functions}(c), the convergence of the focusing function to the focal location is ensured by the first arrival, whereas the coda of the focusing function removes unwanted reflections caused by the first arrival when it encounters reflectors while propagating to the focal location. The arrival times of the direct wave of the focusing function are the same as the arrival times of the direct wave of the time-reversed Green's function. The difference is that the coda of the focusing function is designed to cancel out the events that are introduced by the direct arrival, whereas the coda of the time-reversed Green's function introduces additional artifacts, in the form of reverberations. \\
\subsection{Marchenko method}
We use the Marchenko method to retrieve the focusing functions and Green's functions required for the representations for homogeneous Green's function retrieval. A more detailed consideration of the method can be found in \citet{wapenaar2014marchenko}. Here we only consider the equations and properties of the method relevant for this paper. The Green's function and focusing function of a medium are related via the reflection response:
\begin{equation} \label{T10}
G(\mathbf{x},\mathbf{x}_B,t) - F(\mathbf{x}_B,\mathbf{x},-t)=\int_{\partial\mathbb{D}_0}\int_{-\infty}^\infty R(\mathbf{x},\mathbf{x}_S,t')F(\mathbf{x}_B,\mathbf{x}_S,t-t')\text{d}t'\text{d}^2\mathbf{x}_S,
\end{equation}
where $\mathbf{x}_B$ is a location inside the medium and $\mathbf{x}_S$ indicates the array of sources that are present on the non-enclosing surface $\partial\mathbb{D}_0$. \Eqnref{T10} states that if the reflection response $R$ at a boundary $\partial\mathbb{D}_0$ and a focusing function with a focal location inside medium $\mathbb{D}$ are available, the Green's function with a source at the focal location can be retrieved. The retrieval of the focusing function inside the medium can be achieved using the iterative Marchenko equation:
\begin{equation} \label{A5}
\begin{split}
F_{k+1}(\mathbf{x}_B,\mathbf{x},-t)&=D(\mathbf{x},\mathbf{x}_B,t)\\
&-w(\textbf{x},\textbf{x}_B,t)\int_{\partial\mathbb{D}_0}\int_{-\infty}^\infty R(\mathbf{x},\mathbf{x}_S,t')F_k(\mathbf{x}_B,\mathbf{x}_S,t-t')\text{d}t'\text{d}^2\mathbf{x}_S,
\end{split}
\end{equation}
where $F_{k}(\mathbf{x}_B,\mathbf{x},t)$ is the estimated focusing function after $k$ iterations,  $D(\mathbf{x},\mathbf{x}_B,t)$ is the first arrival of the Green's function and $w(\textbf{x},\textbf{x}_B,t)$ is a windowing function. The windowing function is used to mute the Green's function completely. When the windowing function is applied to \eqnref{T10}, the Green's function is removed. The arrival times of the first arrival of the Green's function are the same as the arrival times of the last arrival of the time-reversed focusing function, hence the windowing function also removes the last arrival of the time-reversed focusing function, however, it will not remove the coda of the time-reversed focusing function. Therefore, in order to obtain the full focusing function in \eqnref{A5}, $D(\mathbf{x},\mathbf{x}_B,t)$ needs to be added after the windowing function has been applied. The windowing function $w(\textbf{x},\textbf{x}_B,t)$ can be estimated from $D(\mathbf{x},\mathbf{x}_B,t)$, as the edge of the muting area is located around the first arrival. In order to use \eqnref{A5} and start the iterative scheme, a first estimation of the focusing function is required. The time-reversed first arrival $D(\mathbf{x}_S,\mathbf{x}_B,-t)$ of the Green's function is used as this first estimation of the focusing function $F_0(\mathbf{x}_B,\mathbf{x}_S,t)$. If this arrival is emitted into the medium, it will cause additional reflections that are not cancelled. By using \eqnref{A5} iteratively, until convergence is achieved, the coda of the focusing function is retrieved, which will suppress the undesired reflections. The only required components for the iterative scheme are a reflection response measured at the single-sided boundary (i.e. the Earth's surface) and the direct arrival from the focal point to the same boundary. This direct arrival can be modeled using a smooth velocity model. Because only the direct arrival is of interest, the model requires no detailed features. Generally, a monopole point source is used to model these first arrivals. After the focusing function has been retrieved, it can be used in \eqnref{T10} to compute the Green's function, $G(\mathbf{x},\mathbf{x}_B,t)$. Subsequently, this Green's function and a similarly derived focusing function for focal point $\textbf{x}_A$ are used in \eqnref{T6} to retrieve the homogeneous Green's function, $G_{\rm h}(\mathbf{x}_A,\mathbf{x}_B,\omega)$. All the Green's functions and focusing functions in this paper are retrieved using the Marchenko method to ensure the representations are applied to the field data and the synthetic data in the same way.\\
The Marchenko method has restrictions, particularly when it is applied on field data. An important underlying assumption of the Marchenko method that is considered in this paper, is that no free-surface multiples are present in the reflection response. Hence, the free-surface multiples should be removed prior to applying the Marchenko method, for example by applying  a surface-related multiple elimination scheme \citep{verschuur1992adaptive}. There are ways to incorporate these multiples in the Marchenko method as well, for an example, see \citet{singh2015marchenko}. Additionally, the reflection response that is used needs to be accurate, as issues with the quality of the recording have strong influences on the final result. An important requirement is that the medium of interest needs to be lossless, which is an unrealistic approximation in real media. Also, the reflection response needs, preferably, to be densely sampled, contain both positive and negative source-receiver offsets and have sufficient recording length and aperture. The effects of some of the limitations of the Marchenko method are considered in \cite{ravasi2016target}, \citet{brackenhoff2016rescaling} and \citet{staring2017adaptive}. When synthetic data are used, the reflection response can be modeled without these limitations. However, when field data are recorded, not all of these requirements are fulfilled and appropriate pre-processing is required.
\subsection{Double-couple source}
As mentioned before, the Green's functions and focusing functions that are retrieved using the Marchenko method usually have a monopole source signature. In the field, especially for passive recordings, there are many different types of source mechanisms. These different source mechanisms can be taken into account for the homogeneous Green's function retrieval as well. We show this by incorporating a double-couple source mechanism in the representation. This mechanism is chosen because it is considered to be representative for an earthquake reponse when the wavelength of the wavefield is larger than the dimensions of the source \citep{aki2002quantitative}. The mechanism differs from a monopole source in the sense that the radiation pattern is not homogeneous, rather, the polarity and amplitude vary depending on the radiation angle of the source. There will be four distinct polarity changes along the radiaton pattern, at 90, 180, 270 and 360 degrees. The source can be inclined at various angles, for example aligned along a fault, which changes the orientation of the polarity changes as well.\\ 
Operator $\mathfrak{D}_B\{\cdot\}$ is introduced, which transforms a monopole source signature to a double-couple source signature. This operator is defined as
\begin{equation} \label{DCop}
\mathfrak{D}_B\{\cdot\} = (\theta_i^{\parallel}+\theta_i^{\perp})\partial_{i,B}\{\cdot\},
\end{equation}
where $\partial_{i,B}$ is a component of the vector containing the partial derivatives acting on the monopole signal originating from source location $\mathbf{x}_B$, to transform the source signature to a double-couple mechanism, $\theta_i^{\parallel}$ is a component of the unit vector that orients one couple of the signal parallel to the fault plane and $\theta_i^{\perp}$ is a component of a similar vector that orients the other couple perpendicular to the fault plane. The operator is applied to the representation for homogeneous Green's function retrieval in \eqnref{T6}, which yields:
\begin{equation} \label{DC2}
\mathfrak{D}_B\{G_{\rm h}(\mathbf{x}_A,\mathbf{x}_B,\omega)\} =  \mathfrak{I}\int_{\partial\mathbb{D}_0}\frac{4}{\omega\rho(\mathbf{x})}\{ \partial_3F(\mathbf{x}_A,\mathbf{x},\omega)\mathfrak{D}_B\{G(\mathbf{x},\mathbf{x}_B,\omega)\}\}\text{d}^2\mathbf{x},
\end{equation}
where the subscript $B$ in the double-couple operator indicates that the operator acts on the source in location $\mathbf{x}_B$. In this representation, the double-couple operator is only applied to the Green's function for the virtual source location, $G(\mathbf{x},\mathbf{x}_B,\omega)$, and not to the focusing function, $F(\mathbf{x}_A,\mathbf{x},\omega)$, for the virtual receiver location. This is because the focusing function is retrieved using \eqnref{A5}, and thus has a monopole source signature. For the retrieval of the Green's function, $\mathfrak{D}_B\{G(\mathbf{x},\mathbf{x}_B,t)\}$, we apply the double-couple operator to \eqnrefs{T10} and \eqref{A5}:
\begin{equation} \label{DC3}
\begin{split}
\mathfrak{D}_B\{G(\mathbf{x},\mathbf{x}_B,t)\} &- \mathfrak{D}_B\{F(\mathbf{x}_B,\mathbf{x},-t)\}\\
&=\int_{\partial\mathbb{D}_0}\int_{-\infty}^\infty R(\mathbf{x},\mathbf{x}_S,t')\mathfrak{D}_B\{F(\mathbf{x}_B,\mathbf{x}_S,t-t')\}\text{d}t'\text{d}^2\mathbf{x}_S,
\end{split}
\end{equation}
\begin{equation} \label{DC4}
\begin{split}
\mathfrak{D}_B\{F_{k+1}(\mathbf{x}_B,\mathbf{x},-t)\}&=\mathfrak{D}_B\{D(\mathbf{x},\mathbf{x}_B,-t)\}-w(\textbf{x},\textbf{x}_B,t)\\
&\times\int_{\partial\mathbb{D}_0}\int_{-\infty}^\infty R(\mathbf{x},\mathbf{x}_S,t')\mathfrak{D}_B\{F_k(\mathbf{x}_B,\mathbf{x}_S,t-t')\}\text{d}t'\text{d}^2\mathbf{x}_S.
\end{split}
\end{equation}
The first arrival is modeled using the double-couple source mechanism instead of a monopole source mechanism, and used in \eqnref{DC4}, for $k$=$0$, as $\mathfrak{D}_B\{F_0(\mathbf{x}_B,\mathbf{x}_S,t)\}$. This results in an estimation of the focusing function $\mathfrak{D}_B\{F(\mathbf{x}_B,\mathbf{x},t)\}$, with a double-couple source mechanism. This focusing function is then used in \eqnref{DC3} to obtain the Green's function $\mathfrak{D}_B\{G(\mathbf{x},\mathbf{x}_B,t)\}$ with a double-couple source signature, which in turn is used in \eqnref{DC2}.
\newpage
\section{Datasets}
We consider both a synthetic dataset and a field dataset. The synthetic dataset is based on the field dataset, hence we first consider the parameters of the field recording before the synthetic dataset is considered.
\subsection{V\o ring data}
\begin{figure}[!hptb]
\centering
\includegraphics[width=\columnwidth]{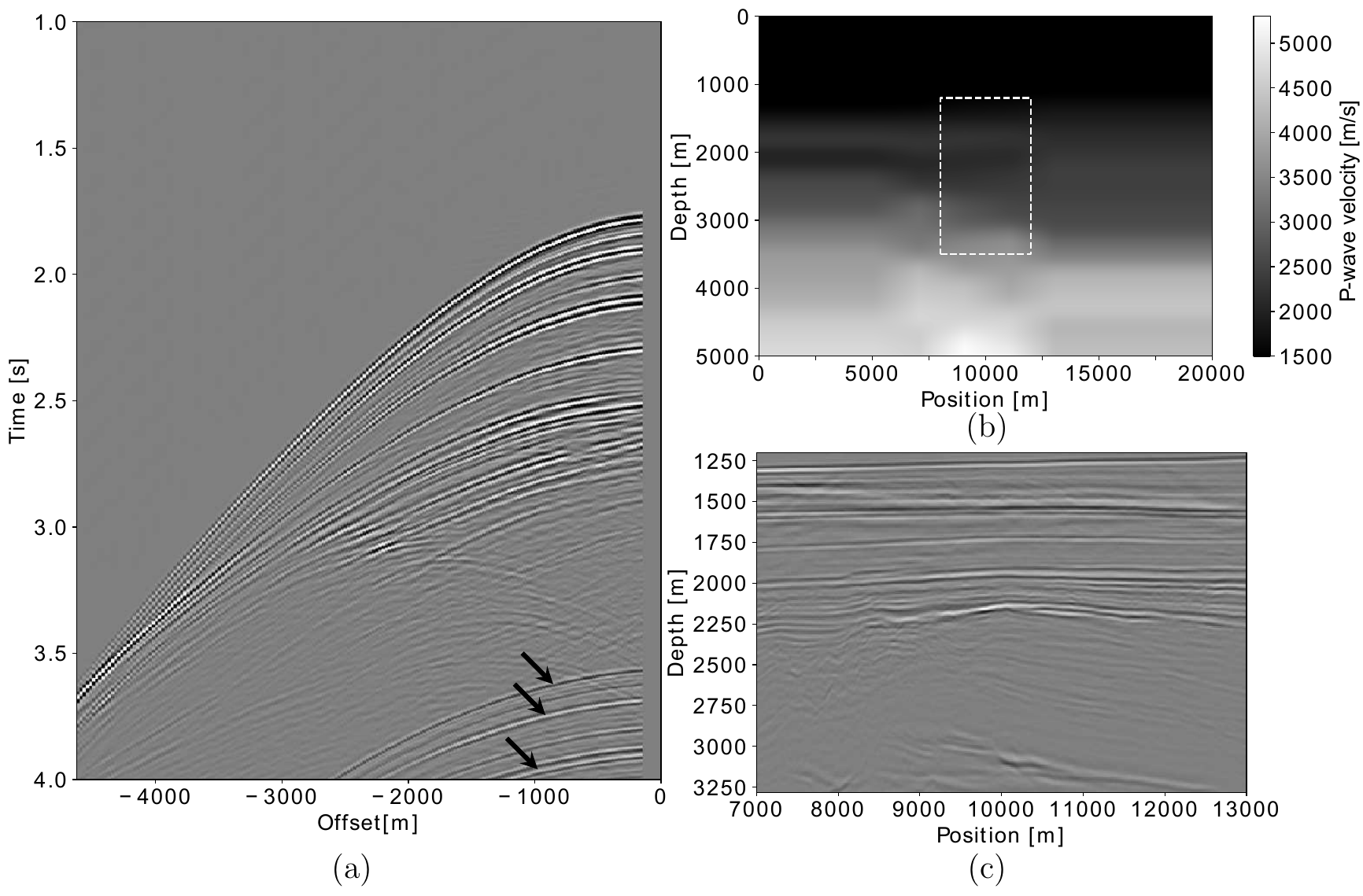}
\caption{(a) \textit{Unprocessed common-source record of the reflection response recorded in the V\o ring basin. The source is located at zero offset, where along with other near offsets no data could be recorded. Free-surface multiples are present in the data at late times, indicated by the black arrows. }(b)\textit{ Estimated P-wave velocity model in $\frac{m}{s}$ of the subsurface region where the data in the V\o ring basin were recorded. The white dashed box represents the region of interest. }(c)\textit{ Image of the region of interest, indicated by the white dashed box in }(b)\textit{. The wavelet in the data in }(a)\textit{ and }(c)\textit{ has been reshaped into a 30 Hz Ricker wavelet for display purposes.}}
\label{saga}
\end{figure}

The considered field data were recorded in a marine setting over the V\o ring basin in offshore Norway by SAGA Petroleum A.S., which is currently part of Equinor. The data consist of a reflection response acquired along a 2D line. For each source location, the receivers were moved along with the source position. The parameters of the acquisition can be found in Table \ref{tab1}. An example of a single common-source record is shown in \figref{saga}(a), where the wavelet on the data has been reshaped to a 30 Hz Ricker wavelet for display purposes (a Ricker wavelet is defined as minus the second time-derivative of a Gaussian function). There are several events present in the common-source record, however it should be noted that the near offsets are missing. This is a common situation for marine acquisition, because receivers cannot be placed too close to active sources. The sources and receivers are located inside the water, and because S-waves cannot propagate in water, only P-waves are measured by the receivers. There are conversions from P-waves to S-waves and back in the subsurface below the water, so there are P-waves present in the reflection data that were converted from S-waves. The data also contain free-surface multiples, which are indicated by the black arrows. \\ 
\begin{table}[!hptb]
\centering
\caption{\textit{Acquisition parameters for the V\o ring dataset.}}
\begin{tabular}{l l}
\hline
Parameter & Value \\ \hline
Number of source positions & 399\\ 
Source spacing & 25 $m$ \\ 
First source position & 5000 $m$ \\ 
Final source position & 14950 $m$ \\ 
Number of receiver positions per source & 180 \\ 
Receiver spacing & 25 $m$ \\ 
Minimum source-receiver offset & 150 $m$ \\ 
Maximum source-receiver offset & 4625 $m$ \\ 
Number of recording samples & 2001 \\ 
Sampling interval & 0.004 $s$ \\ 
High-cut frequency & 90 Hz \\ \hline
\end{tabular}
\label{tab1}
\end{table}
Aside from the reflection data, a smooth P-wave velocity model is provided by Equinor and displayed in \figref{saga}(b). This model is used to create the first arrivals required for the Marchenko method. The dashed white box in \figref{saga}(b) indicates the region of interest where the homogeneous Green's functions are retrieved in this paper. An image of the region of interest was constructed, which is shown in \figref{saga}(c), using the reflection data, the velocity model and a one-way recursive depth migration based on \citet{thorbecke2004design}. Imaging is not the main subject of this paper, so the details of the construction of this image are not discussed, however it should be noted that free-surface multiples were removed prior to imaging. Also note that the retrieval of the homogeneous Green's function (discussed in the sections on synthetic data and field data) and the construction of the image were done independently of each other. The image is used to construct a subsurface model and to validate the homogeneous Green's function retrieval on the field data.  \\
\subsection{Synthetic data}
\begin{figure}[!hptb]
\centering
\includegraphics[width=\columnwidth]{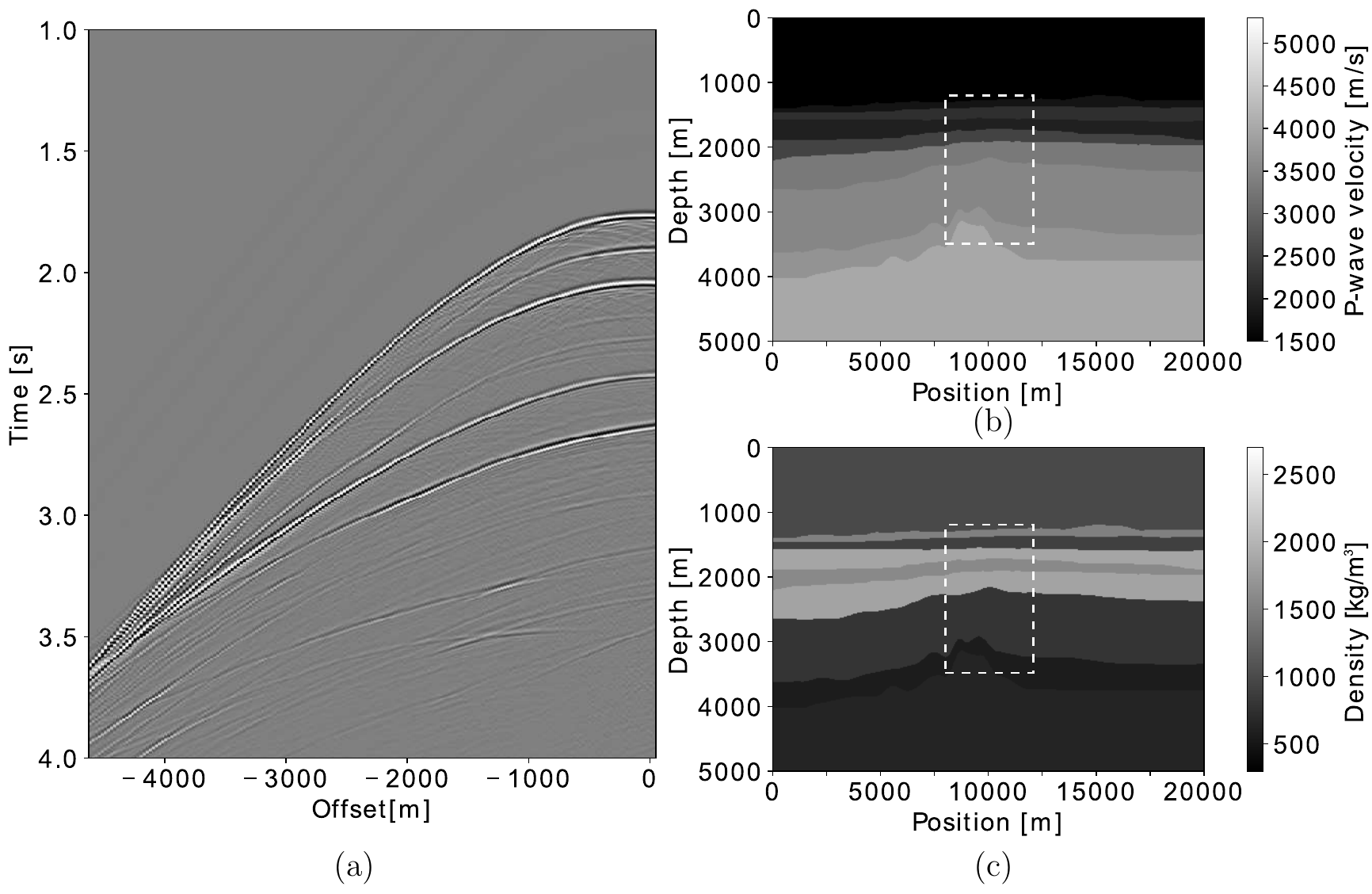}
\caption{(a)\textit{ Common-source record of the acoustic reflection response, modeled using finite-difference modeling in the P-wave velocity model from }(b)\textit{ and density model from }(c)\textit{, at the same location as the common-source record in \figref{saga}}(a)\textit{. The data have been convolved with a 30 Hz Ricker wavelet for display purposes. }(b)\textit{ Synthetic P-wave velocity model in $\frac{m}{s}$ based on the smooth velocity model from \figref{saga}}(b)\textit{ and the image from \figref{saga}}(c)\textit{. }(c)\textit{ Synthetic density model in $\frac{kg}{m^3}$ based on the image from \figref{saga}}(c)\textit{. The white boxes in }(b)\textit{ and }(c)\textit{ indicate the region of interest.}}
\label{synthetic}
\end{figure}

The synthetic models that we use in this paper are based on the field dataset. Because there are no direct measurements of the subsurface available, an acoustic model is interpreted based on the smooth P-wave velocity model in \figref{saga}(b) and the image in \figref{saga}(c). This is done, because there is no S-wave velocity information of the subsurface available, hence we cannot construct an elastodynamic model. It is possible to use an elastodynamic representation for homogeneous Green's function retrieval, see \citet{reinicke2019}, however for the V\o ring basin we do not have the required multi-component data to do so.\\
 To construct the model, we use the image to determine the locations of geological layer boundaries and determine the P-wave velocities by calculating the interval velocities in the smooth P-wave velocity model between the contrasts. The interpreted velocity model is displayed in \figref{synthetic}(b), which shows hard boundaries. Notice that below the area of interest there are no reflectors present in the medium. It is not possible to achieve reliable imaging in this area, therefore no structures are interpreted. Features outside the region of interest were extrapolated to create a full model. A density model is also constructed in order to ensure strong amplitudes in the reflection data. Because no direct measurements of density in the subsurface are available, it is not possible to create a density model that represents the subsurface accurately. Instead, the densities are chosen based on realistic ranges, that ensure strong contrasts between the layers. \Figref{synthetic}(c) displays the density model.\\ 
The finite-difference wavefield modeling code by \citet{thorbecke2011finite} is used to create the single-sided reflection data as input for the Marchenko method. The reflection response of the interpreted model is modeled using the same measurement parameters as for the real dataset, shown in Table \ref{tab1}, with pressure receivers recording the modeled wavefield at the surface of the model, in response to force sources that are located at the same surface. However, all offsets are included, no absorption is modeled and free-surface multiples are ignored. When comparing \figref{saga}(a) and \figref{synthetic}(a), there are similar events, however fewer events are present in the synthetic data. Not all the reflectors in the subsurface can be properly imaged and interpreted, therefore only the major features are present. The field data is more complex than the synthetic data. Because the exact P-wave velocity and density information of the actual medium are not available, there is an amplitude mismatch. The converted waves due to elastic interactions from the actual recording are also not taken into account. Furthermore, no free-surface multiples are present in the modeled reflection data.
\newpage
\section{Synthetic data}
\begin{figure}[!htb]
\centering
\includegraphics[width=1.0\columnwidth]{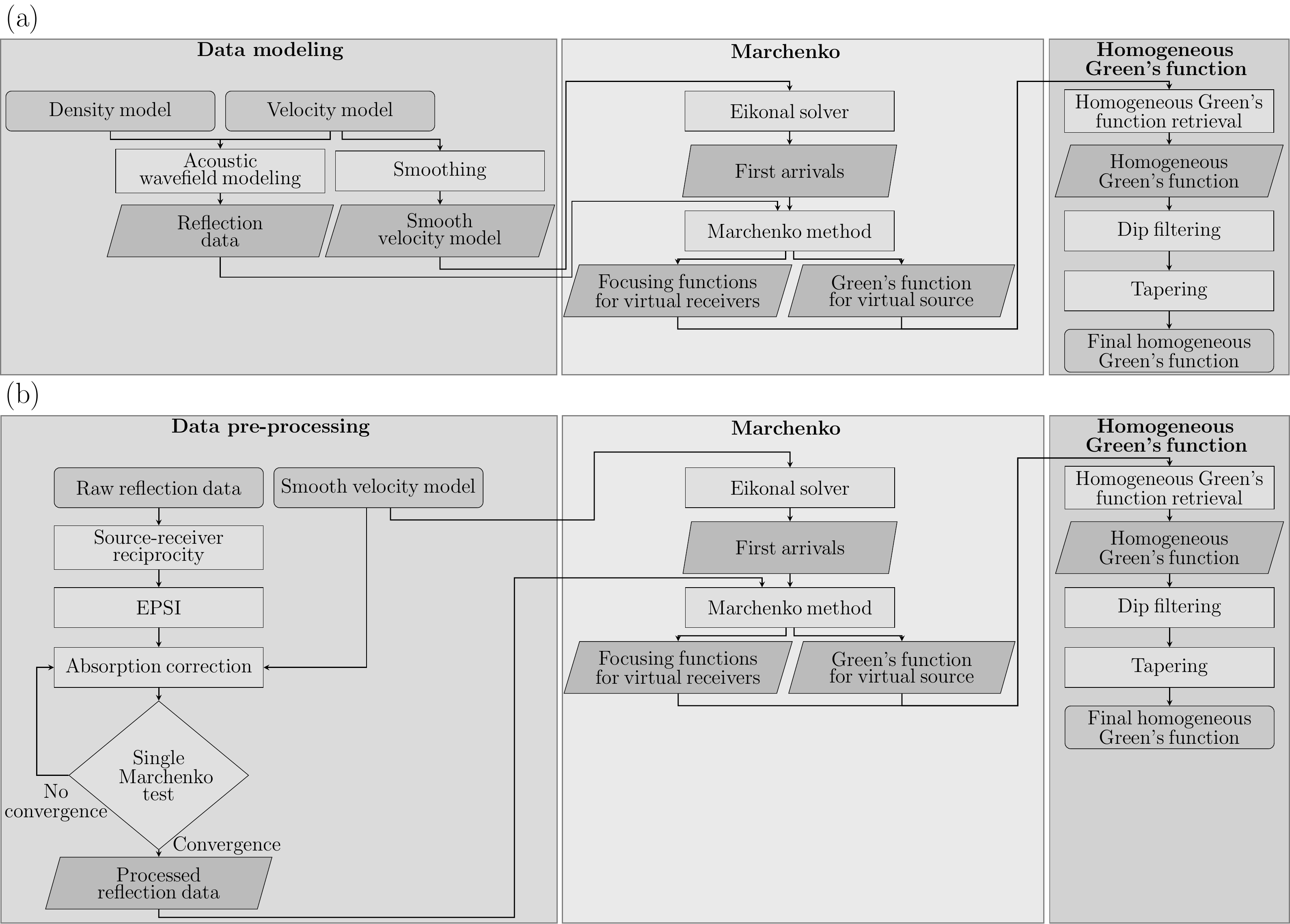}
\caption{\textit{Flowchart for the retrieval of the homogeneous Green's function for (a) the synthetic data and (b) the field data.}}
\label{flowchart}
\end{figure}
The synthetic data are used to retrieve the homogeneous Green's function. The single-sided reflection data and the velocity model are the only information employed. Furthermore, the velocity model is smoothed to create a background velocity model, without any detailed features. This smooth velocity model is used in an eikonal solver that is based on the method by \citet{vidale1988finite} to determine the arrival times from every location in the region of interest, that is indicated in \figrefs{synthetic}(b) and (c), to the single-sided surface. These arrival times are combined with amplitude estimations using the method by \citet{spetzler2005ray} and convolved with a wavelet to create a band-limited pressure wavefield with a monopole source signature. Note that only the smooth velocity model is used for this approach and no density information is used. The first arrivals and the reflection data are used as input for the iterative Marchenko equation, as described by \eqnref{A5}, using the code by \citet{thorbecke2017implementation}. The windowing function is created by using the arrival times from the eikonal solver. Through use of the iterative code, Green's functions and focusing functions are obtained.
\subsection{Homogeneous Green's function retrieval}
\begin{figure}[!hptb]
\centering
\includegraphics[width=1.0\columnwidth]{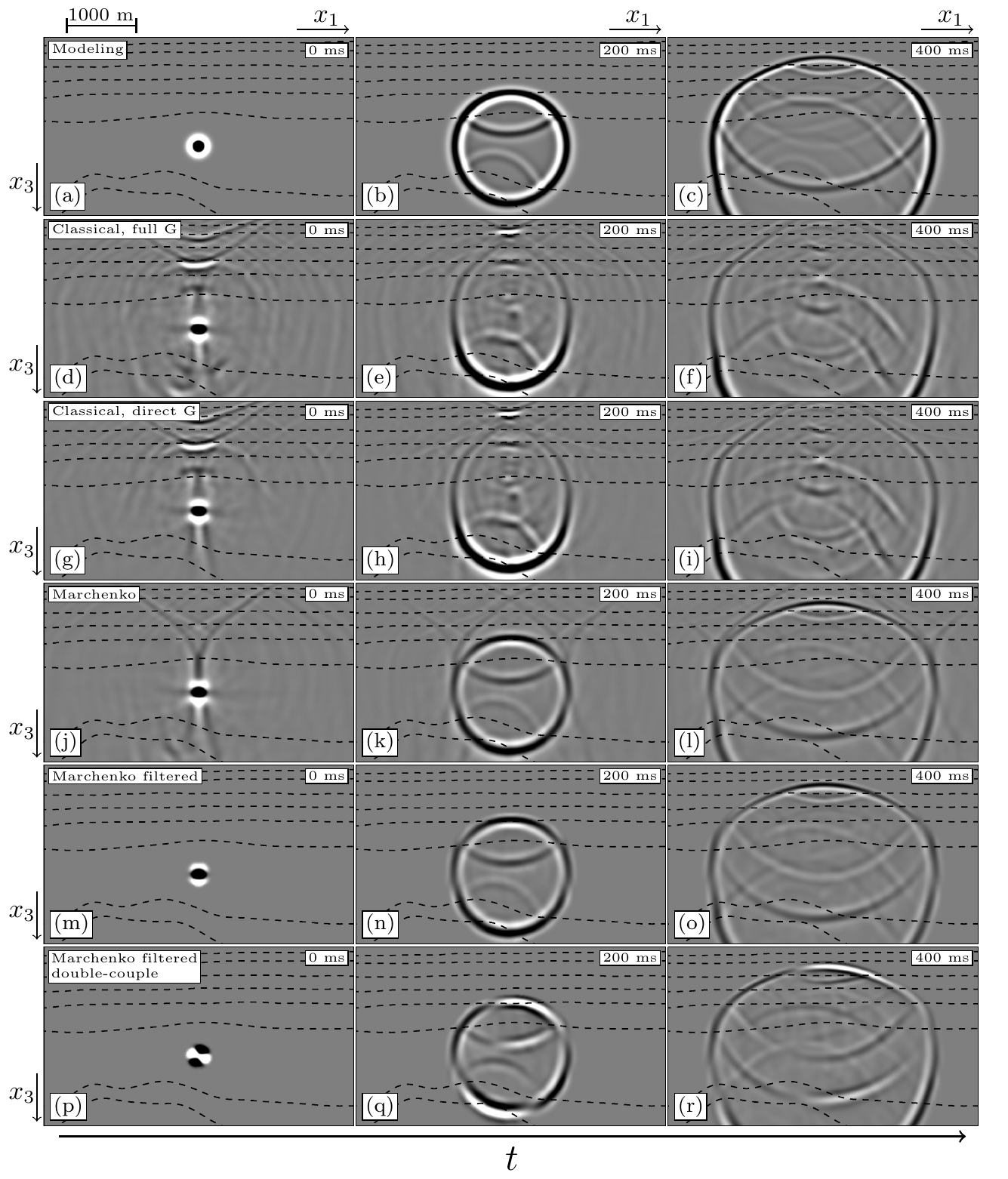}
\vspace{-1.0cm}
\caption{{\setstretch{1.0}\textit{Snapshots of the wavefield at different times. Column 1 indicates t=0 ms, column 2 t=200 ms and column 3 t=400 ms. All wavefields have been convolved with a 15 Hz Ricker wavelet for display purposes. The black dotted lines indicate the locations of geological layer interfaces.} (a)-(c) \textit{Directly modeled homogeneous Green's function in the subsurface used as a reference.} (d)-(f) \textit{Homogeneous Green's function retrieval using \eqnref{T5} with full Green's functions for the virtual source and receiver positions.} (g)-(i) \textit{Idem, however with a full Green's function for the virtual receiver position and a direct arrival as the Green's function for the virtual source position.} (j)-(l) \textit{Homogeneous Green's function retrieval using \eqnref{T6}, a Green's function for the virtual source position and a focusing function for the virtual receiver position, without filtering.} (m)-(o) \textit{Idem, with dip filtering and tapering applied.} (p)-(r) \textit{Idem, using a Green's function with a double-couple source signature inclined at 20 degrees.}}}
\label{homgm}
\end{figure}

To demonstrate the advantage of the single-sided representation, we compare different ways of retrieving the homogeneous Green's function. We aim to replicate a realistic situation as much as possible, hence the only data that are used are the smooth velocity model and the modeled reflection response. First, to create a benchmark for the homogeneous Green's function, we model the wavefield directly inside the medium by placing a volume injection-rate source and pressure receivers inside the region of interest. The wavefield is time-reversed and added to the original wavefield conforming to \eqnref{T21} to create the homogeneous Green's function. Three snapshots of this result are shown in \figrefs{homgm}(a)-(c) at 0 ms, 200 ms and 400 ms, respectively. This is the result to which the homogeneous Green's functions that are obtained using the representations in \eqnrefs{T5} and \eqref{T6} are compared to. In \figref{homgm}, the wavefield is convolved with a 15 Hz Ricker wavelet and dotted black lines are shown for reference where we expect scattering to take place. This type of visualization is used for all the snapshots we produce for the synthetic data.\\ 
Now, we assume that we do not know the exact model and use the Marchenko method to retrieve Green's functions in the region of interest in the subsurface, from the reflection data at the single-sided surface at the top of the model. These Green's functions are used to evaluate \eqnref{T5} at the single-sided surface, not at an enclosing surface. The location $\textbf{x}_B$ of one response, $G(\mathbf{x},\mathbf{x}_B,t)$, is kept constant as the virtual source position, while the location $\textbf{x}_A$ of the other response, $G(\mathbf{x}_A,\mathbf{x},t)$, varies to serve as the virtual receiver positions. The positions of the virtual source and receivers are exactly the same as the respective source and receiver positions of the modeled wavefield. The retrieved homogeneous Green's function is shown in \figrefs{homgm}(d)-(f) at 0 ms, 200 ms and 400 ms, respectively. In \figref{homgm}(d), for zero time, there is noise present in the entire snapshot, and while there is a focus of the wavefield to the source position, there are strong artifacts present directly above the source position. Furthermore, coherent artefacts with weaker amplitudes are present throughout the snapshot. The snapshot at 200 ms in \figref{homgm}(e) shows that the downgoing primary wavefield is retrieved, however, the upgoing primary wavefield is missing, instead forming into a weakening event with incorrect arrival times. Furthermore, while the coda of the wavefield has been retrieved with the correct upgoing reflections, the downgoing reflections are missing, and downgoing artifacts from above the source location are present in the snapshot. Aside from these problems, the snapshot is contaminated by artifacts. The final snapshot at 400 ms in \figref{homgm}(f) shows similar problems. \\
For comparsion, we repeat the experiment, however, instead of using a full Green's function $G(\mathbf{x},\mathbf{x}_B,t)$ for the virtual source, we only use its first arrival to reduce the number of artifacts. The results are shown in \figrefs{homgm}(g)-(i) at 0 ms, 200 ms and 400 ms, respectively. Compared to the previous experiment, the number of artifacts decreases, although not all are removed. The strong source artifacts at time zero remain present and the upgoing primary wavefield and downgoing coda are not restored.\\
Next we apply the single-sided representation using \eqnref{T6}. For the virtual source location $\mathbf{x}_S$ the same Green's function, $G(\mathbf{x},\mathbf{x}_B,t)$, is used as in the previous two experiments. However, the Green's function that was used to create the virtual receiver, $G(\mathbf{x}_A,\mathbf{x},t)$, is replaced by a focusing function, $F(\mathbf{x}_A,\mathbf{x},t)$. \Figrefs{homgm}(j)-(l) shows the result at 0 ms, 200 ms and 400 ms, respectively. The improvement is noticeable, as artifacts are removed from the homogeneous Green's function. Aside from this, the primary wavefield is fully reconstructed, as is the coda of the downgoing wavefield. When comparing this result to the benchmark, it shows that the events are retrieved at the correct locations and times, although an amplitude mismatch is present. This is due to the fact that the amplitude of the first arrival is not exact, because we assume that we cannot model the first arrival in the real medium. When the amplitude of the first arrival is incorrect it scales the retrieved focusing functions and Green's function, although the correct relative amplitude, arrival times and events are obtained \citep{brackenhoff2016rescaling,brackenhoff2019}. Some of the events are not reconstructed, especially when the angle of the reflection is high. This is because the single-sided boundary is assumed to be infinite, while in reality the aperture is limited. The reflection response lacks certain angles of reflection, so these angles cannot be reconstructed for the homogeneous Green's function. At zero time the snapshot contains less artifacts, however, some remain, which contaminate the homogeneous Green's function at later times.\\
An analysis of the wavenumber-frequency spectrum of the homogeneous Green's function retrieved using \eqnref{T6} revealed noise at locations corresponding to high angles of the wavefield. As mentioned before, the aperture of the reflection response is limited, so at these angles no events can be reconstructed. However, this also introduces noise into the final result as the Marchenko method tries to reconstruct these angles. To remove the noise at these high angles, dip filtering is applied to the homogeneous Green's function, which can be applied because no physical events are reconstructed for these angles. This creates some small artifacts around the first arrival of the homogeneous Green's function, which in turn are removed by applying a time-dependent taper. The improved result is shown in \figref{homgm}(m)-(o) at 0 ms, 200 ms and 400 ms, respectively. None of the desired events have been removed, however, the artifacts around the source position are gone. When this result is compared to the modeled response, we find the match of the arrival times quite satisfactory, while the amplitude mismatch remains, and the improvement over the homogeneous Green's function obtained using the classical representation is significant. When comparing the result of this retrieval to the modeled result from \figref{homgm}(a)-(c), the match between desired events and their arrival times are strong. A notable difference occurs in the shape of the source at zero time. When the wavefield is modeled directly, the source radiates uniformly in all directions, however, when the wavefield is retrieved with the Marchenko method, this is not possible. This is caused by the limited aperture, which cannot capture the parts of the wavefield that propagate horizontally at this depth. Additionally, the dip filtering that is applied will also affect the source radiation. Even with these limitations, this type of workflow yields the optimal result, and is used for the retrieval of the homogeneous Green's function from here onward. The workflow is summarized in \figref{flowchart}(a).\\
Finally, we consider the workflow in \figref{flowchart}(a) with a double-couple source mechanism. Instead of using \eqnref{T6} for the homogeneous Green's function retrieval, we use \eqnref{DC2}. This means that we use a Green's function, with a double-couple source signature, obtained with the Marchenko method, to investigate whether the source signature has a strong effect on the retrieved homogeneous Green's function. A first arrival caused by a double-couple source, needed to initiate the Marchenko method, cannot be retrieved using the eikonal solver, therefore we model the required first arrival using the finite-difference code. The double-couple source mechanism is incorporated into the code using the moment tensor approach for finite-difference modeling, as described by \citet{li2014global}. The double-couple source mechanism is an elastic mechanism, however, it is assumed that all the data are acoustic. To model an acoustic response to the double-couple source, we use the smoothed P-wave velocity model and a homogeneous density model of 1000 $\frac{{\rm kg}}{{\rm m^3}}$, respectively, as well as a homogenous S-wave velocity model of 1000 $\frac{{\rm m}}{{\rm s}}$, except for the top layer where the S-wave velocity is set to zero, simulating a marine setting. This means that no S-waves will arrive at the receiver locations. The coda of this modeling will be incorrect, however, as we only use the first arrival, this has no consequences for our results, aside from the scaling of the first arrival. The first arrival will be a pure P-wave, as the velocities of the P-wave model are larger than those of the S-wave model, which is a realistic situation.\\ 
We use a double-couple source that was inclined at -20 degrees to model the first arrival that it is used in the Marchenko method to create a Green's function. This Green's function replaces the one created from a monopole source that was used in the previous examples and the resulting homogeneous Green's function is shown in \figrefs{homgm}(p)-(r). The arrival times of the events are the same as the ones in the previous experiment where only monopole signatures were employed. The main differences are found in the polarity of the events. Due to the double-couple source, the amplitude and polarity of the wavefield changes depending on the angle. Because the source we used to model the first arrival for the Green's function is inclined at -20 degrees, these polarity changes are not occuring at 90, 180, 270 and 360 degrees, but, shifted by -20 degrees,  at approximately 70, 160, 250 and 340 degrees. All retrieved events, not just the first arrival, are affected by this inclination, without introducing any additional artifacts. This shows that different source signatures can be incorporated into the single-sided representation for homogeneous Green's function retrieval.
\subsection{Limitations of reflection data}
\begin{figure}[!hptb]
\centering
\includegraphics[width=1.0\columnwidth]{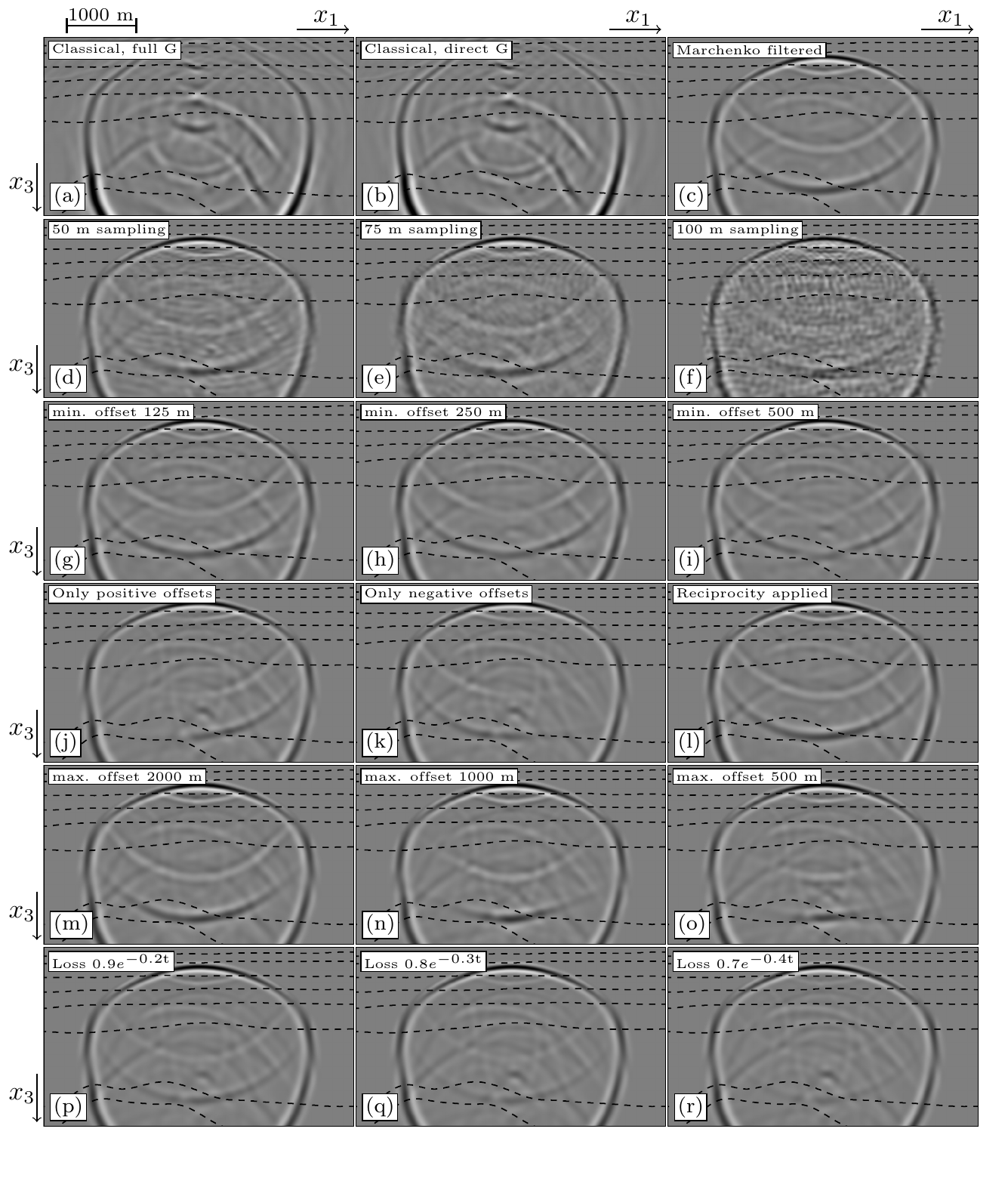}
\vspace{-1.8cm}
\caption{\small \textit{Snapshots of the homogeneous Green's function at t=400 ms, retrieved using varying limitations of the reflection response, following the workflow in \figref{flowchart}(a). All wavefields have been convolved with a 15 Hz Ricker wavelet.} (a) \textit{Result from \figref{homgm}}(f), (b) \textit{result from \figref{homgm}}(i) \textit{and} (c) \textit{result from \figref{homgm}}(o) \textit{, the latter used as a benchmark for the other results. Second row: Homogeneous Green's function obtained, when the reflection response has a source and receiver spacing of} (d) \textit{50 m,} (e) \textit{75 m and }(f) \textit{100 m. }\textit{Third row: Homogeneous Green's function obtained when the reflection response is missing the small source-receiver offsets up to a distance of} (g) \textit{125 m,} (h) \textit{250 m and }(i) \textit{500m.} \textit{Fourth row: Homogeneous Green's function obtained when the reflection response contains} (j) \textit{only positive source-receiver offsets,} (k) \textit{only negative source-receiver offsets and }(l) \textit{has all source-receiver  offsets restored using source-receiver reciprocity.} \textit{Fifth row: Homogeneous Green's function obtained when the reflection response has an aperture limited to} (m) \textit{2000 m,} (n) \textit{1000 m and }(o) \textit{500m.} \textit{Sixth row: Homogeneous Green's function obtained when the reflection response has a loss applied to it of} (p) \textit{$0.9e^{-0.2t}$,} (q) \textit{$0.8e^{-0.3t}$ and }(r) \textit{$0.7e^{-0.4t}$.}}
\label{homglim}
\end{figure}

The reflection response that we use to retrieve the results in \figref{homgm} is nearly ideal, due to the recording setup and the absence of absorption. In the following experiment, we perform the homogeneous Green's function retrieval using the workflow in \figref{flowchart}(a), with five different types of acquisition limitations applied to the reflection response. In all five cases we perform the entire process that is outlined in the workflow, starting with the Marchenko method to retrieve the focusing function and Green's function from the reflection response, to which one of the limitations is applied, followed by applying \eqnref{T6} to obtain the homogeneous Green's function. This demonstrates the effects of the limitations of the reflection response on the homogeneous Green's function that is obtained. The results of these tests are shown in \figref{homglim}, where (a), (b) and (c), show the result from \figref{homgm}(f), (i) and (o), respectively, which are used as a reference. All snapshots in \figref{homglim}, are shown at 400 ms. The results shown in the rows below the first one are achieved in the same way as the result from \figref{homglim}(c), with different types of limitations applied to the reflection response. Each column shows a varying value of the limitation to indicate the sensitivity of the method to these limitations.\\ 
To determine the Nyquist spatial sampling interval for the data, we use the sampling criterion: 
\begin{equation} \label{samp}
\Delta x < \frac{c_1}{2f_{max}sin(\theta_{max})},
\end{equation}
where $\Delta x$ is the spatial sampling, $c_1$ is the velocity of the top layer of the subsurface, $f_{max}$ is the maximum frequency of the wavelet and $\theta_{max}$ is the maximum angle of the waves in the top layer. To determine the sampling, we use the top layer velocity of the data, which corresponds to 1500 $\frac{\rm m}{\rm s}$, a "maximum" frequency of 30 Hz for the 15 Hz Ricker wavelet and a maximum angle of 60 degrees. This results in a recommended spatial sampling of about 29 meters. The current sampling of 25 meters should therefore be sufficient for our applications, which is supported by the quality of the retrieved homogeneous Green's function that have been obtained in the previous section. \\
In \figrefs{homglim}(d)-(f), we display the result retrieved using a reflection response that has a coarse source-receiver sampling. The sampling values for the receiver and source spacing are 50, 75 and 100 meters, which are double, triple and quadruple the original spacing, and all exceed the Nyquist sampling. When the spacing is doubled, noise is introduced into the homogeneous Green's function due to spatial aliasing. The physical events are distorted by this noise and background artifacts are present. This issue is worsened when the spacing distance is tripled. Some events are obscured and strong noise is present. Quadrupling the spacing produces a result that is unusable. It consists almost entirely of noise and the physical events cannot be distinguished. Hence, for successful use of the method the source-receiver sampling of the reflection response must not be larger than the Nyquist sampling. \\
Next, we consider the influence of missing small source-receiver offsets. The result is shown in \figrefs{homglim}(g)-(i), where, respectively, the first 125, 250 and 500 meters of the offsets are removed from the reflection response, for both positive and negative offsets, and replaced with empty traces. When 125 meters of offsets are missing, the homogeneous Green's function is still comparable to the ideal situation. There is a degradation in quality and a few artifacts are present. Removing 250 meters of near offsets produces a less accurate homogeneous Green's function, with a stronger degradation in quality. When 500 meters of near offsets are removed, the low angle reflections are not reconstructed properly, similarly to the high angle noise that is caused by the limited aperture of the data. The events below and above the virtual source position are missing and strong artifacts are present. These artifacts can be partially removed by adjusting the dip filtering to remove the low angle events, however, this procedure will also remove part of the physical events and is therefore not reccomendable to apply. The near offsets do have an impact on the final result and ideally should be reconstructed, if possible, before applying the Marchenko method.\\
Aside from missing near offsets, a reflection response can also be recorded exclusively in one direction. \Figrefs{homglim}(j) shows the result using only positive source-receiver offsets in the reflection response and \figref{homglim}(k) does the same for negative source-receiver offsets. In both cases unwanted artifacts are present and, depending on the direction of the source-receiver offsets, large parts of the physical events are missing. This issue can be avoided by applying source-receiver reciprocity. We perform source-receiver reciprocity on the reflection response containing only the negative offsets and retrieve the result shown in \figref{homglim}(l). This homogeneous Green's function is similar to the one produced in the ideal situation. We retrieve a similar result when we apply source-receiver reciprocity on the reflection response containing only positive offsets.\\
The final acquisition limitation that we review is the absence of large source-receiver offsets, or aperture limitation of the data. In \figrefs{homglim}(m)-(o), we show the homogeneous Green's function when the largest source-receiver offset is, respectively, 2000, 1000 and 500 meters. When the aperture is 2000 meters, the homogeneous Green's function is comparable to the one that is retrieved in the ideal situation, with some artifacts introduced. When the aperture is limited to 1000 meters, a homogeneous Green's function is produced that contains more artifacts and is missing a larger portion of the angles of the desired events. This is once again due to the fact that the angles of this part of the wavefield are not present in the reflection response. If only 500 meters of aperture are available, a large portion of the angles of events are missing. This is clear when \figref{homglim}(o) is compared to \figref{homglim}(i). The parts of the events that are missing due to the lack of near offsets are present in the case of limited aperture and vice versa. By applying a stronger dip filter, the artifacts can be suppressed, however, as mentioned before, this also removes part of the physical events.\\
Finally, we consider the case of absorption, which is a factor that cannot directly be influenced during the acquistion of the reflection response. Even if the recording setup is ideal, absorption of the wavefield is present and can degrade the result. This is demonstrated in \figrefs{homglim}(p)-(r), where the loss is simulated by applying time-dependent absorption functions to the reflection data of $0.9e^{-0.2{\rm t}}$, $0.8e^{-0.3{\rm t}}$ and $0.7e^{-0.4{\rm t}}$, respectively. In case of low absorption, the homogeneous Green's function still contains the physical events, although they have a lower amplitude. The artifacts are present with a low amplitude. If the absorption is increased, the physical events start to vanish and the artifacts are more pronounced. In case of high absorption, the physical events have very low amplitude and there are strong artifacts present.
\newpage
\section{Field Data}
\subsection{Pre-processing}
\begin{figure}[!hptb]
\centering
\includegraphics[width=\columnwidth]{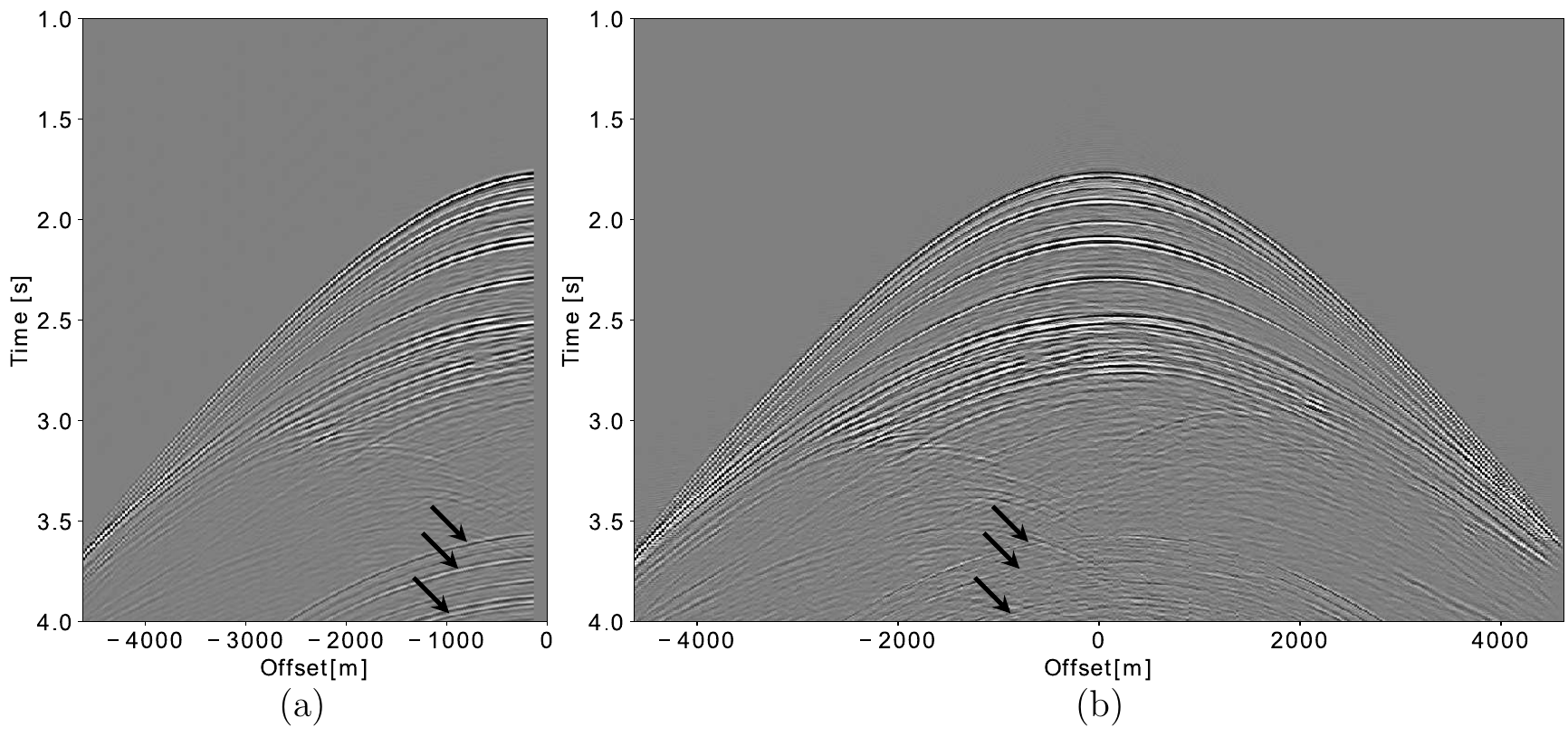}
\caption{(a) \textit{Common-source record from \figref{saga}}(a) \textit{before any processing is applied and }(b)\textit{ common-source record from} (a) \textit{with source-receiver reciprocity, EPSI and an exponential gain of $1.73e^{1.3t}$ applied. Both common-source records have their wavelets reshaped to a 30 Hz Ricker wavelet. Note the removal of the free-surface multiples at late times, as indicated by the black arrows.}}
\label{shots}
\end{figure}

The raw seismic field reflection data cannot directly be used with the Marchenko method, because when the method uses these data it does not converge to a solution. The reflection data needs to be preprocessed to compensate for the limitations \citep{staring2017adaptive}. As was shown by the results in \figref{homglim}, there are multiple effects that need to be taken into account. The reflection data are missing small source-receiver offsets, only have offsets in a negative direction, may be sub-sampled, possibly have insufficient aperture and are affected by absorption. Aside from this, the 2D line was recorded in a 3D setting, which can cause complications due to out of plane effects.\\ 
The effects of having geometric spreading in a 3D setting while the Marchenko scheme that is applied is 2D, are corrected by applying a time-dependent gain on the data. This gain is equal to $\sqrt{t}$, as the geometrical spreading in 3D can be approximated as being $\frac{1}{t}$ and in 2D as $\frac{1}{\sqrt{t}}$ \citep{berk}. Next, source-receiver reciprocity is applied to the reflection data to create offsets in the positive direction. Having offsets in both positive and negative directions is vital for the next step, where the Estimation of Primaries through Sparse Inversion (EPSI) method is applied. The EPSI method estimates the primaries in the data, through the use of an inversion process, and estimates the free-surface multiples separately. This allows for the retrieval of a datset without the free-surface multiples. Furthermore, in this process, the information about the subsurface that is contained in the free-surface multiples is used to reconstruct the missing near offsets of the primaries. Simultaneously, an estimation of the source wavelet is made, which can be used to deconvolve the reflection data. Note, that the EPSI method only removes the free-surface multiples. The internal multiples remain in the data after the application of the EPSI method. See \citet{van2009estimating} for a detailed overview of the EPSI method.\\ 
The effects of absorption on the data is adaptively corrected for by applying exponential time-gain and a scaling factor. Note that the exponential time-gain is applied in addition to the $\sqrt{t}$ gain that was applied to account for the geometrical spreading factor. This second gain is intended to counteract the effects of the absorption on the data and therefore is estimated separately from the first gain. The first estimations of the time-gain are based on the velocity model conforming to the method discussed by \citet{draganov2010seismic}. The scaling factor is estimated by minimizing the cost functions proposed by \citet{brackenhoff2016rescaling}. After these processing steps are applied, the reflection data are used to retrieve a Green's function for one location in the subsurface. If the method does not converge to a solution, where the artifacts are minimal, the exponential gain and scaling factor are adjusted and the test is run again. After a few iterations, we found that applying a gain of $1.73e^{1.3{\rm t}}$ to the data resulted in a solution that converged with significant removal of artifacts.\\ 
Interpolation was also tested by interpolating the source-receiver spacing on the reflection data to smaller values, however we found that this did not significantly improve our results. Another aspect we cannot improve is the limited aperture of the data. The final workflow for the field data can be found in \figref{flowchart}(b). An example of a common-source record before and after the pre-processing is shown in \figref{shots}. Note the improvement of the common-source record, as the coverage of the data has been increased and the free-surface multiples have been weakened, as indicated by the black arrows. \\
\subsection{Homogeneous Green's function retrieval}
\begin{figure}[!hptb]
\centering
\includegraphics[width=1.0\columnwidth]{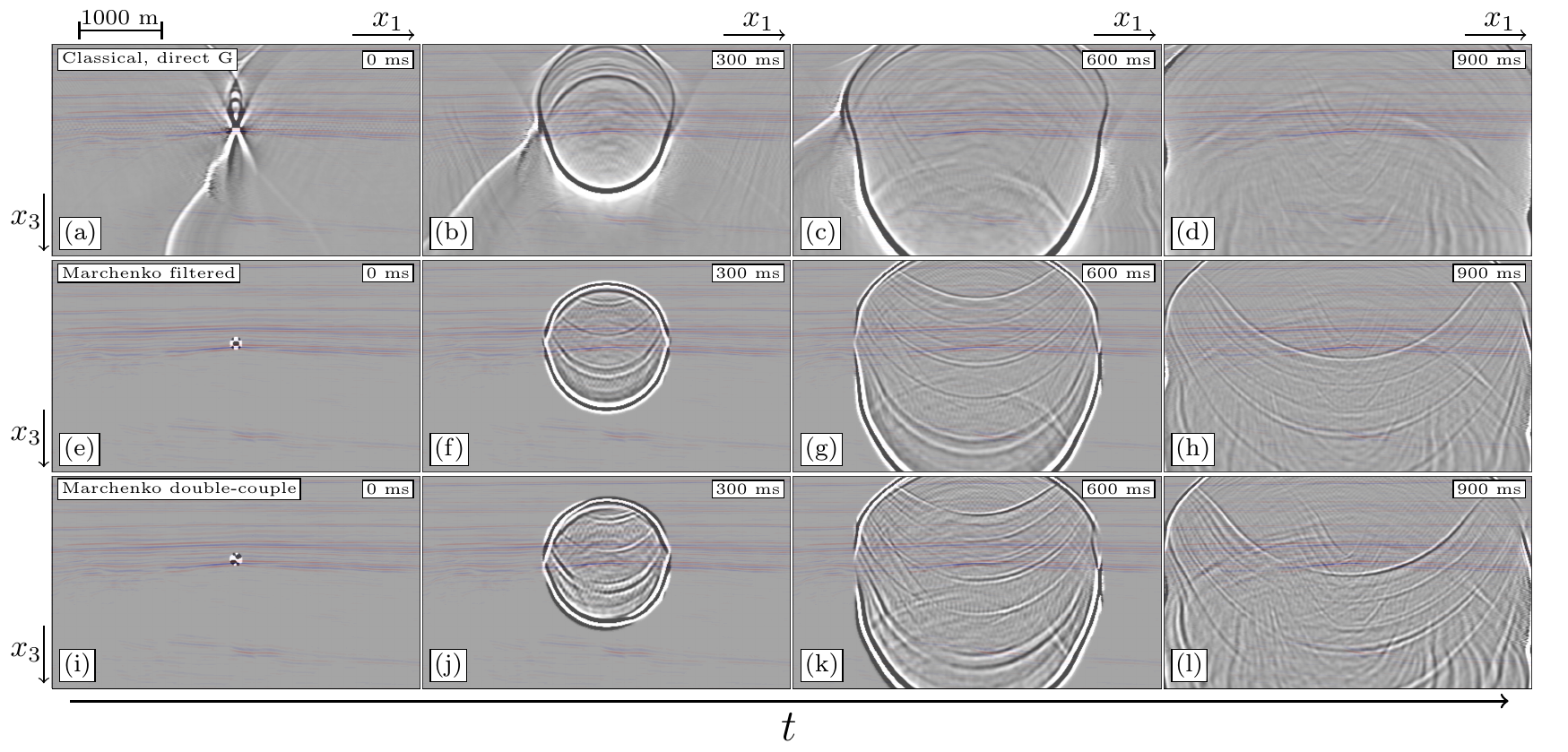}
\caption{\textit{Snapshots of the homogeneous Green's function in the subsurface of the V\o ring basin, obtained using the classical representation of \eqnref{T5} for a monopole source at} (a) \textit{0 ms,} (b) \textit{300 ms,} (c) \textit{600 ms and} (d) \textit{900 ms. Idem, using the single-sided representation of \eqnref{T6} for a monopole source at} (e) \textit{0 ms,} (f) \textit{300 ms,} (g) \textit{600 ms and} (h) \textit{900 ms. Idem, using the single-sided representation of \eqnref{DC2} for a double-couple source inclined at -20 degrees at} (i) \textit{0 ms,} (j) \textit{300 ms,} (k) \textit{600 ms and} (l) \textit{900 ms. All data contain a 30 Hz Ricker wavelet.}}
\label{homgfield}
\end{figure}

After applying all the corrections, the Marchenko method is utilized to retrieve the required Green's functions and focusing functions in the region of interest that is indicated in \figref{saga}(b). These functions are retrieved using only the pre-processed single-sided reflection response and a smooth velocity model. Next, we use these functions to retrieve the homogeneous Green's function using the single-sided boundary representation of \eqnref{T6}. Snapshots of the obtained result are shown in \figrefs{homgfield}(e)-(h), for 0, 300, 600 and 900 ms, respectively. For comparison, the homogeneous Green's function obtained using the classical representation and using only the first arrival for the Green's function for the virtual source is shown in \figrefs{homgfield}(a)-(d), for 0, 300, 600 and 900 ms, respectively. In both cases a monopole source was used to model all the first arrivals.\\ 
We also test the creation of a virtual source that has a double-couple source signature on the field data. We use a double-couple source inclined at -20 degrees to create the first arrival that is used to obtain the Green's function for the virtual source using the Marchenko method. Similar to our approach on the synthetic data, we use the smooth P-wave velocity model, a homogeneous density model of 1000 $\frac{{\rm kg}}{{\rm m^3}}$ and a constant S-wave velocity model of 1000 $\frac{{\rm m}}{{\rm s}}$, with the top layer of water, where the S-wave velocity is zero. We select the P-wave first arrival and use the Marchenko method to create the Green's function for the virtual source. The homogeneous Green's function is obtained through \eqnref{DC2} and the workflow in \figref{flowchart}(b). Snapshots of the result are shown in \figref{homgfield}(i)-(l) for 0, 300, 600 and 900 ms, respectively.\\ 
For all images, a transparent overlay of the image from \figref{saga}(c) is used to indicate locations where scattering is expected. This image is only used for verification and was not used for the retrieval of the homogeneous Green's function. The results of the single-sided representation for the monopole source from \figref{homgfield}(e)-(h) were previously shown in \citet{wapenaar2018virtual}, and the results of the single-sided representation for the double-couple source from \figref{homgfield}(i)-(l) were previously shown in \citet{brackenhoff2019}. The details about the pre-processing of the reflection data were not discussed before. \\
The snapshots of the homogeneous Green's functions obtained using the single-sided representation for both types of sources show multiple events, both upgoing and downgoing. The locations of the scattering and the layer interfaces in the image overlay have a strong match and, aside from the primary reflections, the multiple reflections can also be seen. All of these events are completely absent when the classical representation is used. Strong artifacts are present in this case and the coda of the downgoing wavefield is missing entirely. The primary downgoing wavefield is present, however the upgoing primary wavefield is absent, which is similar to the result that was obtained on the synthetic data. This shows that, as expected, the single-sided representation is an improvement over the classical representation on field data as well. The results for the double-couple source signature are similar to the ones that were obtained on the synthetic data. The polarity changes are present on both the primary wavefield as well as on the coda. Because of these changes there are a few events in the coda where the amplitude is low, which makes it harder to distinguish these events. The results in \figref{homgfield} are different from those that were obtained in \figref{homgm}. As stated in the Datasets section, this is because the synthetic data do not contain all the events that are present in the field data. The results on the field data contain more events, which could not be predicted from the image. This is an advantage of the data-driven approach of the Marchenko method. We find the overall results in \figref{homgfield} encouraging. \\ 
However, even the homogeneous Green's functions obtained using the single-sided representation are not perfect as there are still artifacts present. This is partially due to the presence of background noise in the dataset, which distorts the final result. More coherent events are also present, which do not correlate with the primary wavefield and scattering locations from the image. Because we cannot be sure whether the reflection response has been pre-processed perfectly, there may be some low amplitude artifacts present that are nor fully compensated for or which are created by the Marchenko method. The Marchenko method that we applied was intended for 2D acoustic media, however, the true medium is 3D and elastic. As the geological layering for this region appears to be close to horizontal, the out-of-plane effects are assumed to be small. Due to conversion from P-waves to S-waves and back, some events are present in the reflection response that would not be present if the medium was purely acoustic and these are not handled correctly by our acoustic Marchenko implementation. \\
\subsection{Discussion}
The results on the field data show promise for further applications of the single-sided representation, for example, by combining them with passive measurements. In this case, the Green's function for the virtual source, that was constructed through the use of the Marchenko method, would be replaced with a passive recording. The accurately retrieved propagation and scattering of the wavefield in the inhomogeneous medium also holds much promise, despite the presence of some artifacts. If we wanted to track the paths of the wavefronts emitted for example by an induced seismic source, a result like this could provide substantial insight. However, there are several limitations. In this paper, we exclusively considered a 2D dataset that has few out-of-plane effects. For complex 3D media, where the out-of-plane effects are much more severe, 3D reflection data would be required, with sufficient coverage, as well as a 3D smooth P-wave velocity model. In this case, the issues with offset, sampling and aperture that were present for the 2D data would be more complex to deal with, and will be present in two directions. Furthermore, passive measurement setups are usually sparser than for active measurements, which could yield additional complications if one wants to use a passive recording for the source of the homogeneous Green's function. These aspects are subject of ongoing research.
\newpage
\section{Conclusion}
We demonstrated the data-driven generation of virtual receivers and a virtual source, which have the potential to improve, for example, the monitoring of the subsurface and the prediction of the complex response of different source mechanisms. We did this by utilizing a single-sided representation to retrieve the homogeneous Green's function in the subsurface. To this end, we applied the Marchenko method, which only requires a single-sided reflection response and a smooth velocity model. We showed that the single-sided representation produces significantly more accurate results than the classical representation when the reflection data are only available on a single-sided non-enclosing boundary, typically the Earth's surface. The sensitivity of the Marchenko method to limitations of the reflection data was investigated by manipulating synthetic reflection data. This showed that pre-processing of the reflection response to compensate for coarse source-receiver sampling, missing offsets and absorption is vital for the succesful application of the Marchenko method.\\
We considered a dataset from the V\o ring basin (offshore Norway), which was affected by some of such limitations and processed the data using geometric spreading correction, source-receiver reciprocity, the EPSI method and applying a time-gain and scaling factor. The processed reflection response was used to obtain the focussing functions and Green's functions, needed to apply the representations for homogeneous Green's function retrieval. The homogeneous Green's function obtained using the single-sided representation shows potential in our opinion for wavefield monitoring in the subsurface, as the complete coda of the wavefield is recovered, which is not the case when the classical representation is used. The scattering occurs at locations that correlate with possible reflector locations. Monopole source and double-couple source signatures can both be used in the Marchenko method and in the single-sided representation to obtain homogeneous Green's functions with the same source signature. To further explore this potential, more complex source mechanisms should be considered, such as dynamic fault planes, that are active over an extended area and time period. This includes taking into account the effects caused by elastic media instead of acoustic media.

\acknowledgments
The authors thank Jan-Willem Vrolijk and Eric Verschuur of the Delphi Consortium for help with processing the data and Equinor for giving permission to use the vintage seismic reflection data of the V\o ring Basin. The authors also wish to thank the anonymous reviewers for their constructive comments. This project has received funding from the European Research Council (ERC) under the European Union's Horizon 2020 research and innovation programme (grant agreement No: 742703). The authors declare no competing interests. The stable versions of the finite-difference modeling code and Marchenko code used in this paper can be found on \url{https://janth.home.xs4all.nl/Software/OpenSource.tgz} \citep{jan_thorbecke_2019_3374728}, while the versions in progress can be found on \url{https://github.com/JanThorbecke/OpenSource}. Additionally, for some of the figures, the SUpython package was used, which can be found on \url{https://github.com/Jbrackenhoff/SUpython}. The seismic reflection data analysed are available from Equinor, but restrictions apply to the availability of these data, which were used under license for the current study, and so are not publicly available. Data are however available from the authors upon reasonable request and with permission of Equinor.

\newpage
\bibliography{JGR}

\begin{thebibliography}{39}
\expandafter\ifx\csname natexlab\endcsname\relax\def\natexlab#1{#1}\fi
\expandafter\ifx\csname bibnamefont\endcsname\relax
  \def\bibnamefont#1{#1}\fi
\expandafter\ifx\csname bibfnamefont\endcsname\relax
  \def\bibfnamefont#1{#1}\fi
\expandafter\ifx\csname citenamefont\endcsname\relax
  \def\citenamefont#1{#1}\fi
\expandafter\ifx\csname url\endcsname\relax
  \def\url#1{\texttt{#1}}\fi
\expandafter\ifx\csname urlprefix\endcsname\relax\def\urlprefix{URL }\fi
\providecommand{\bibinfo}[2]{#2}
\providecommand{\eprint}[2][]{\url{#2}}

\bibitem[{\citenamefont{Yilmaz}(2001)}]{yilmaz2001seismic}
\bibinfo{author}{\bibfnamefont{{\"O}.}~\bibnamefont{Yilmaz}},
  \emph{\bibinfo{title}{Seismic data analysis}}, vol.~\bibinfo{volume}{1}
  (\bibinfo{publisher}{Society of exploration geophysicists Tulsa, OK},
  \bibinfo{year}{2001}).

\bibitem[{\citenamefont{Grigoli et~al.}(2017)}]{grigoli2017current}
\bibinfo{author}{\bibfnamefont{F.}~\bibnamefont{Grigoli}} \bibnamefont{et~al.},
  \bibinfo{journal}{Reviews of Geophysics} \textbf{\bibinfo{volume}{55 (2)}},
  \bibinfo{pages}{310} (\bibinfo{year}{2017}).

\bibitem[{\citenamefont{van Thienen-Visser and
  Breunese}(2015)}]{van2015induced}
\bibinfo{author}{\bibfnamefont{K.}~\bibnamefont{van Thienen-Visser}}
  \bibnamefont{and} \bibinfo{author}{\bibfnamefont{J.}~\bibnamefont{Breunese}},
  \bibinfo{journal}{The Leading Edge} \textbf{\bibinfo{volume}{34 (6)}},
  \bibinfo{pages}{664} (\bibinfo{year}{2015}).

\bibitem[{\citenamefont{Magnani et~al.}(2017)\citenamefont{Magnani, Blanpied,
  DeShon, and Hornbach}}]{magnani2017discriminating}
\bibinfo{author}{\bibfnamefont{M.~B.} \bibnamefont{Magnani}},
  \bibinfo{author}{\bibfnamefont{M.~L.} \bibnamefont{Blanpied}},
  \bibinfo{author}{\bibfnamefont{H.~R.} \bibnamefont{DeShon}},
  \bibnamefont{and} \bibinfo{author}{\bibfnamefont{M.~J.}
  \bibnamefont{Hornbach}}, \bibinfo{journal}{Science Advances}
  \textbf{\bibinfo{volume}{3 (11)}}, \bibinfo{pages}{e1701593}
  (\bibinfo{year}{2017}).

\bibitem[{\citenamefont{Porter}(1970)}]{porter1970diffraction}
\bibinfo{author}{\bibfnamefont{R.~P.} \bibnamefont{Porter}},
  \bibinfo{journal}{Journal of the Optical Society of America}
  \textbf{\bibinfo{volume}{60 (8)}}, \bibinfo{pages}{1051}
  (\bibinfo{year}{1970}).

\bibitem[{\citenamefont{Porter and Devaney}(1982)}]{porter1982holography}
\bibinfo{author}{\bibfnamefont{R.~P.} \bibnamefont{Porter}} \bibnamefont{and}
  \bibinfo{author}{\bibfnamefont{A.~J.} \bibnamefont{Devaney}},
  \bibinfo{journal}{Journal of the Optical Society of America}
  \textbf{\bibinfo{volume}{72 (3)}}, \bibinfo{pages}{327}
  (\bibinfo{year}{1982}).

\bibitem[{\citenamefont{Oristaglio}(1989)}]{oristaglio1989inverse}
\bibinfo{author}{\bibfnamefont{M.~L.} \bibnamefont{Oristaglio}},
  \bibinfo{journal}{Inverse Problems} \textbf{\bibinfo{volume}{5 (6)}},
  \bibinfo{pages}{1097} (\bibinfo{year}{1989}).

\bibitem[{\citenamefont{Wapenaar}(2004)}]{wapenaar2004retrieving}
\bibinfo{author}{\bibfnamefont{K.}~\bibnamefont{Wapenaar}},
  \bibinfo{journal}{Physical Review Letters} \textbf{\bibinfo{volume}{93}},
  \bibinfo{pages}{254301} (\bibinfo{year}{2004}).

\bibitem[{\citenamefont{van Manen et~al.}(2005)\citenamefont{van Manen,
  Robertsson, and Curtis}}]{van2005modeling}
\bibinfo{author}{\bibfnamefont{D.-J.} \bibnamefont{van Manen}},
  \bibinfo{author}{\bibfnamefont{J.~O.} \bibnamefont{Robertsson}},
  \bibnamefont{and} \bibinfo{author}{\bibfnamefont{A.}~\bibnamefont{Curtis}},
  \bibinfo{journal}{Physical Review Letters} \textbf{\bibinfo{volume}{94
  (16)}}, \bibinfo{pages}{164301} (\bibinfo{year}{2005}).

\bibitem[{\citenamefont{Bakulin and Calvert}(2005)}]{bakulin2005virtual}
\bibinfo{author}{\bibfnamefont{A.}~\bibnamefont{Bakulin}} \bibnamefont{and}
  \bibinfo{author}{\bibfnamefont{R.}~\bibnamefont{Calvert}}, in
  \emph{\bibinfo{booktitle}{{SEG} Technical Program Expanded Abstracts 2005}}
  (\bibinfo{publisher}{Society of Exploration Geophysicists},
  \bibinfo{address}{{Houston}, {Texas}}, \bibinfo{year}{2005}).

\bibitem[{\citenamefont{Curtis et~al.}(2009)\citenamefont{Curtis, Nicolson,
  Halliday, Trampert, and Baptie}}]{curtis2009virtual}
\bibinfo{author}{\bibfnamefont{A.}~\bibnamefont{Curtis}},
  \bibinfo{author}{\bibfnamefont{H.}~\bibnamefont{Nicolson}},
  \bibinfo{author}{\bibfnamefont{D.}~\bibnamefont{Halliday}},
  \bibinfo{author}{\bibfnamefont{J.}~\bibnamefont{Trampert}}, \bibnamefont{and}
  \bibinfo{author}{\bibfnamefont{B.}~\bibnamefont{Baptie}},
  \bibinfo{journal}{Nature Geoscience} \textbf{\bibinfo{volume}{2 (10)}},
  \bibinfo{pages}{700} (\bibinfo{year}{2009}).

\bibitem[{\citenamefont{Wapenaar et~al.}(2017)\citenamefont{Wapenaar,
  Thorbecke, van~der Neut, Slob, and Snieder}}]{wapenaar2017virtual}
\bibinfo{author}{\bibfnamefont{K.}~\bibnamefont{Wapenaar}},
  \bibinfo{author}{\bibfnamefont{J.}~\bibnamefont{Thorbecke}},
  \bibinfo{author}{\bibfnamefont{J.}~\bibnamefont{van~der Neut}},
  \bibinfo{author}{\bibfnamefont{E.}~\bibnamefont{Slob}}, \bibnamefont{and}
  \bibinfo{author}{\bibfnamefont{R.}~\bibnamefont{Snieder}},
  \bibinfo{journal}{Geophysical Prospecting} \textbf{\bibinfo{volume}{65 (6)}},
  \bibinfo{pages}{1430} (\bibinfo{year}{2017}).

\bibitem[{\citenamefont{Wapenaar et~al.}(2016)\citenamefont{Wapenaar,
  Thorbecke, and van~der Neut}}]{wapenaar2016singleGH}
\bibinfo{author}{\bibfnamefont{K.}~\bibnamefont{Wapenaar}},
  \bibinfo{author}{\bibfnamefont{J.}~\bibnamefont{Thorbecke}},
  \bibnamefont{and} \bibinfo{author}{\bibfnamefont{J.}~\bibnamefont{van~der
  Neut}}, \bibinfo{journal}{Geophysical Journal International}
  \textbf{\bibinfo{volume}{205 (1)}}, \bibinfo{pages}{531}
  (\bibinfo{year}{2016}).

\bibitem[{\citenamefont{Broggini et~al.}(2012)\citenamefont{Broggini, Snieder,
  and Wapenaar}}]{broggini2012focusing}
\bibinfo{author}{\bibfnamefont{F.}~\bibnamefont{Broggini}},
  \bibinfo{author}{\bibfnamefont{R.}~\bibnamefont{Snieder}}, \bibnamefont{and}
  \bibinfo{author}{\bibfnamefont{K.}~\bibnamefont{Wapenaar}},
  \bibinfo{journal}{Geophysics} \textbf{\bibinfo{volume}{77 (5)}},
  \bibinfo{pages}{A25} (\bibinfo{year}{2012}).

\bibitem[{\citenamefont{Rose}(2001)}]{rose2001single}
\bibinfo{author}{\bibfnamefont{J.~H.} \bibnamefont{Rose}},
  \bibinfo{journal}{Physical Review A} \textbf{\bibinfo{volume}{65 (1)}},
  \bibinfo{pages}{012707} (\bibinfo{year}{2001}).

\bibitem[{\citenamefont{Wapenaar et~al.}(2013)\citenamefont{Wapenaar, Broggini,
  Slob, and Snieder}}]{wapenaar2013three}
\bibinfo{author}{\bibfnamefont{K.}~\bibnamefont{Wapenaar}},
  \bibinfo{author}{\bibfnamefont{F.}~\bibnamefont{Broggini}},
  \bibinfo{author}{\bibfnamefont{E.}~\bibnamefont{Slob}}, \bibnamefont{and}
  \bibinfo{author}{\bibfnamefont{R.}~\bibnamefont{Snieder}},
  \bibinfo{journal}{Physical Review Letters} \textbf{\bibinfo{volume}{110}},
  \bibinfo{pages}{084301} (\bibinfo{year}{2013}).

\bibitem[{\citenamefont{Wapenaar et~al.}(2014)\citenamefont{Wapenaar,
  Thorbecke, van Der~Neut, Broggini, Slob, and
  Snieder}}]{wapenaar2014marchenko}
\bibinfo{author}{\bibfnamefont{K.}~\bibnamefont{Wapenaar}},
  \bibinfo{author}{\bibfnamefont{J.}~\bibnamefont{Thorbecke}},
  \bibinfo{author}{\bibfnamefont{J.}~\bibnamefont{van Der~Neut}},
  \bibinfo{author}{\bibfnamefont{F.}~\bibnamefont{Broggini}},
  \bibinfo{author}{\bibfnamefont{E.}~\bibnamefont{Slob}}, \bibnamefont{and}
  \bibinfo{author}{\bibfnamefont{R.}~\bibnamefont{Snieder}},
  \bibinfo{journal}{Geophysics} \textbf{\bibinfo{volume}{79 (3)}},
  \bibinfo{pages}{WA39} (\bibinfo{year}{2014}).

\bibitem[{\citenamefont{Ravasi et~al.}(2016)\citenamefont{Ravasi, Vasconcelos,
  Kritski, Curtis, da~Costa~Filho, and Meles}}]{ravasi2016target}
\bibinfo{author}{\bibfnamefont{M.}~\bibnamefont{Ravasi}},
  \bibinfo{author}{\bibfnamefont{I.}~\bibnamefont{Vasconcelos}},
  \bibinfo{author}{\bibfnamefont{A.}~\bibnamefont{Kritski}},
  \bibinfo{author}{\bibfnamefont{A.}~\bibnamefont{Curtis}},
  \bibinfo{author}{\bibfnamefont{C.~A.} \bibnamefont{da~Costa~Filho}},
  \bibnamefont{and} \bibinfo{author}{\bibfnamefont{G.~A.} \bibnamefont{Meles}},
  \bibinfo{journal}{Geophysical Journal International}
  \textbf{\bibinfo{volume}{205 (1)}}, \bibinfo{pages}{99}
  (\bibinfo{year}{2016}).

\bibitem[{\citenamefont{Staring et~al.}(2018)\citenamefont{Staring, Pereira,
  Douma, van~der Neut, and Wapenaar}}]{staring2017adaptive}
\bibinfo{author}{\bibfnamefont{M.}~\bibnamefont{Staring}},
  \bibinfo{author}{\bibfnamefont{R.}~\bibnamefont{Pereira}},
  \bibinfo{author}{\bibfnamefont{H.}~\bibnamefont{Douma}},
  \bibinfo{author}{\bibfnamefont{J.}~\bibnamefont{van~der Neut}},
  \bibnamefont{and} \bibinfo{author}{\bibfnamefont{K.}~\bibnamefont{Wapenaar}},
  \bibinfo{journal}{Geophysics} \textbf{\bibinfo{volume}{83 (6)}},
  \bibinfo{pages}{1} (\bibinfo{year}{2018}).

\bibitem[{\citenamefont{Wapenaar et~al.}(2018)\citenamefont{Wapenaar,
  Brackenhoff, Thorbecke, van~der Neut, Slob, and
  Verschuur}}]{wapenaar2018virtual}
\bibinfo{author}{\bibfnamefont{K.}~\bibnamefont{Wapenaar}},
  \bibinfo{author}{\bibfnamefont{J.}~\bibnamefont{Brackenhoff}},
  \bibinfo{author}{\bibfnamefont{J.}~\bibnamefont{Thorbecke}},
  \bibinfo{author}{\bibfnamefont{J.}~\bibnamefont{van~der Neut}},
  \bibinfo{author}{\bibfnamefont{E.}~\bibnamefont{Slob}}, \bibnamefont{and}
  \bibinfo{author}{\bibfnamefont{E.}~\bibnamefont{Verschuur}},
  \bibinfo{journal}{Scientific Reports} \textbf{\bibinfo{volume}{8 (1)}},
  \bibinfo{pages}{2497} (\bibinfo{year}{2018}).

\bibitem[{\citenamefont{Brackenhoff et~al.}(2019)\citenamefont{Brackenhoff,
  Thorbecke, and Wapenaar}}]{brackenhoff2019}
\bibinfo{author}{\bibfnamefont{J.}~\bibnamefont{Brackenhoff}},
  \bibinfo{author}{\bibfnamefont{J.}~\bibnamefont{Thorbecke}},
  \bibnamefont{and} \bibinfo{author}{\bibfnamefont{K.}~\bibnamefont{Wapenaar}},
  \bibinfo{journal}{Solid Earth Discussions} \textbf{\bibinfo{volume}{2019}},
  \bibinfo{pages}{1} (\bibinfo{year}{2019}),
  \urlprefix\url{https://www.solid-earth-discuss.net/se-2018-142/}.

\bibitem[{\citenamefont{Aki and Richards}(2002)}]{aki2002quantitative}
\bibinfo{author}{\bibfnamefont{K.}~\bibnamefont{Aki}} \bibnamefont{and}
  \bibinfo{author}{\bibfnamefont{P.~G.} \bibnamefont{Richards}},
  \emph{\bibinfo{title}{Quantitative seismology}} (\bibinfo{publisher}{W.H.
  Freeman and Company, San Fransisco}, \bibinfo{year}{2002}).

\bibitem[{\citenamefont{Feynman et~al.}(2011)\citenamefont{Feynman, Leighton,
  and Sands}}]{feynman2011feynman}
\bibinfo{author}{\bibfnamefont{R.~P.} \bibnamefont{Feynman}},
  \bibinfo{author}{\bibfnamefont{R.~B.} \bibnamefont{Leighton}},
  \bibnamefont{and} \bibinfo{author}{\bibfnamefont{M.}~\bibnamefont{Sands}},
  \emph{\bibinfo{title}{The {F}eynman lectures on physics, {V}ol. {I}: The new
  millennium edition: mainly mechanics, radiation, and heat}},
  vol.~\bibinfo{volume}{1} (\bibinfo{publisher}{Basic books},
  \bibinfo{year}{2011}).

\bibitem[{\citenamefont{Morse and Feshbach}(1953)}]{Morse53Book}
\bibinfo{author}{\bibfnamefont{P.~M.} \bibnamefont{Morse}} \bibnamefont{and}
  \bibinfo{author}{\bibfnamefont{H.}~\bibnamefont{Feshbach}},
  \emph{\bibinfo{title}{Methods of theoretical physics, {V}ol. {I}}}
  (\bibinfo{publisher}{Mc{G}raw-{H}ill {B}ook {C}ompany {I}nc., {N}ew {Y}ork},
  \bibinfo{year}{1953}).

\bibitem[{\citenamefont{Slob et~al.}(2014)\citenamefont{Slob, Wapenaar,
  Broggini, and Snieder}}]{slob2014seismic}
\bibinfo{author}{\bibfnamefont{E.}~\bibnamefont{Slob}},
  \bibinfo{author}{\bibfnamefont{K.}~\bibnamefont{Wapenaar}},
  \bibinfo{author}{\bibfnamefont{F.}~\bibnamefont{Broggini}}, \bibnamefont{and}
  \bibinfo{author}{\bibfnamefont{R.}~\bibnamefont{Snieder}},
  \bibinfo{journal}{Geophysics} \textbf{\bibinfo{volume}{79 (2)}},
  \bibinfo{pages}{S63} (\bibinfo{year}{2014}).

\bibitem[{\citenamefont{Verschuur et~al.}(1992)\citenamefont{Verschuur,
  Berkhout, and Wapenaar}}]{verschuur1992adaptive}
\bibinfo{author}{\bibfnamefont{D.~J.} \bibnamefont{Verschuur}},
  \bibinfo{author}{\bibfnamefont{A.}~\bibnamefont{Berkhout}}, \bibnamefont{and}
  \bibinfo{author}{\bibfnamefont{C.}~\bibnamefont{Wapenaar}},
  \bibinfo{journal}{Geophysics} \textbf{\bibinfo{volume}{57 (9)}},
  \bibinfo{pages}{1166} (\bibinfo{year}{1992}).

\bibitem[{\citenamefont{Singh et~al.}(2015)\citenamefont{Singh, Snieder,
  Behura, van~der Neut, Wapenaar, and Slob}}]{singh2015marchenko}
\bibinfo{author}{\bibfnamefont{S.}~\bibnamefont{Singh}},
  \bibinfo{author}{\bibfnamefont{R.}~\bibnamefont{Snieder}},
  \bibinfo{author}{\bibfnamefont{J.}~\bibnamefont{Behura}},
  \bibinfo{author}{\bibfnamefont{J.}~\bibnamefont{van~der Neut}},
  \bibinfo{author}{\bibfnamefont{K.}~\bibnamefont{Wapenaar}}, \bibnamefont{and}
  \bibinfo{author}{\bibfnamefont{E.}~\bibnamefont{Slob}},
  \bibinfo{journal}{Geophysics} \textbf{\bibinfo{volume}{80 (5)}},
  \bibinfo{pages}{S165} (\bibinfo{year}{2015}).

\bibitem[{\citenamefont{Brackenhoff}(2016)}]{brackenhoff2016rescaling}
\bibinfo{author}{\bibfnamefont{J.}~\bibnamefont{Brackenhoff}},
  \bibinfo{type}{M.sc. thesis}, \bibinfo{school}{Delft University of
  Technology}, \bibinfo{address}{Delft, Zuid-Holland, the Netherlands}
  (\bibinfo{year}{2016}), \urlprefix\url{repository.tudelft.nl,
  http://resolver.tudelft.nl/uuid:0f0ce3d0-088f-4306-b884-12054c39d5da}.

\bibitem[{\citenamefont{Thorbecke et~al.}(2004)\citenamefont{Thorbecke,
  Wapenaar, and Swinnen}}]{thorbecke2004design}
\bibinfo{author}{\bibfnamefont{J.~W.} \bibnamefont{Thorbecke}},
  \bibinfo{author}{\bibfnamefont{K.}~\bibnamefont{Wapenaar}}, \bibnamefont{and}
  \bibinfo{author}{\bibfnamefont{G.}~\bibnamefont{Swinnen}},
  \bibinfo{journal}{Geophysics} \textbf{\bibinfo{volume}{69}},
  \bibinfo{pages}{1037} (\bibinfo{year}{2004}).

\bibitem[{\citenamefont{Reinicke and Wapenaar}(2019)}]{reinicke2019}
\bibinfo{author}{\bibfnamefont{C.}~\bibnamefont{Reinicke}} \bibnamefont{and}
  \bibinfo{author}{\bibfnamefont{K.}~\bibnamefont{Wapenaar}},
  \bibinfo{journal}{Wave Motion} \textbf{\bibinfo{volume}{89}},
  \bibinfo{pages}{245} (\bibinfo{year}{2019}).

\bibitem[{\citenamefont{Thorbecke and Draganov}(2011)}]{thorbecke2011finite}
\bibinfo{author}{\bibfnamefont{J.}~\bibnamefont{Thorbecke}} \bibnamefont{and}
  \bibinfo{author}{\bibfnamefont{D.}~\bibnamefont{Draganov}},
  \bibinfo{journal}{Geophysics} \textbf{\bibinfo{volume}{76 (6)}},
  \bibinfo{pages}{H1} (\bibinfo{year}{2011}).

\bibitem[{\citenamefont{Vidale}(1988)}]{vidale1988finite}
\bibinfo{author}{\bibfnamefont{J.}~\bibnamefont{Vidale}},
  \bibinfo{journal}{Bulletin of the Seismological Society of America}
  \textbf{\bibinfo{volume}{78 (6)}}, \bibinfo{pages}{2062}
  (\bibinfo{year}{1988}).

\bibitem[{\citenamefont{Spetzler and Angelov}(2005)}]{spetzler2005ray}
\bibinfo{author}{\bibfnamefont{J.}~\bibnamefont{Spetzler}} \bibnamefont{and}
  \bibinfo{author}{\bibfnamefont{P.}~\bibnamefont{Angelov}}, in
  \emph{\bibinfo{booktitle}{75th {SEG} Annual Meeting}}
  (\bibinfo{publisher}{Society of Exploration Geophysicists},
  \bibinfo{address}{{Houston}, {Texas}}, \bibinfo{year}{2005}).

\bibitem[{\citenamefont{Thorbecke et~al.}(2017)\citenamefont{Thorbecke, Slob,
  Brackenhoff, van~der Neut, and Wapenaar}}]{thorbecke2017implementation}
\bibinfo{author}{\bibfnamefont{J.}~\bibnamefont{Thorbecke}},
  \bibinfo{author}{\bibfnamefont{E.}~\bibnamefont{Slob}},
  \bibinfo{author}{\bibfnamefont{J.}~\bibnamefont{Brackenhoff}},
  \bibinfo{author}{\bibfnamefont{J.}~\bibnamefont{van~der Neut}},
  \bibnamefont{and} \bibinfo{author}{\bibfnamefont{K.}~\bibnamefont{Wapenaar}},
  \bibinfo{journal}{Geophysics} \textbf{\bibinfo{volume}{82 (6)}},
  \bibinfo{pages}{WB29} (\bibinfo{year}{2017}).

\bibitem[{\citenamefont{Li et~al.}(2014)\citenamefont{Li, Helmberger, Clayton,
  and Sun}}]{li2014global}
\bibinfo{author}{\bibfnamefont{D.}~\bibnamefont{Li}},
  \bibinfo{author}{\bibfnamefont{D.}~\bibnamefont{Helmberger}},
  \bibinfo{author}{\bibfnamefont{R.~W.} \bibnamefont{Clayton}},
  \bibnamefont{and} \bibinfo{author}{\bibfnamefont{D.}~\bibnamefont{Sun}},
  \bibinfo{journal}{Geophysical Journal International}
  \textbf{\bibinfo{volume}{197 (2)}}, \bibinfo{pages}{1166}
  (\bibinfo{year}{2014}).

\bibitem[{\citenamefont{Berkhout}(1987)}]{berk}
\bibinfo{author}{\bibfnamefont{A.}~\bibnamefont{Berkhout}},
  \emph{\bibinfo{title}{Applied seismic wave theory}}
  (\bibinfo{publisher}{Elsevier, Amsterdam}, \bibinfo{year}{1987}).

\bibitem[{\citenamefont{van Groenestijn and
  Verschuur}(2009)}]{van2009estimating}
\bibinfo{author}{\bibfnamefont{G.}~\bibnamefont{van Groenestijn}}
  \bibnamefont{and}
  \bibinfo{author}{\bibfnamefont{D.}~\bibnamefont{Verschuur}},
  \bibinfo{journal}{Geophysics} \textbf{\bibinfo{volume}{74 (3)}},
  \bibinfo{pages}{A23} (\bibinfo{year}{2009}).

\bibitem[{\citenamefont{Draganov et~al.}(2010)\citenamefont{Draganov, Ghose,
  Ruigrok, Thorbecke, and Wapenaar}}]{draganov2010seismic}
\bibinfo{author}{\bibfnamefont{D.}~\bibnamefont{Draganov}},
  \bibinfo{author}{\bibfnamefont{R.}~\bibnamefont{Ghose}},
  \bibinfo{author}{\bibfnamefont{E.}~\bibnamefont{Ruigrok}},
  \bibinfo{author}{\bibfnamefont{J.}~\bibnamefont{Thorbecke}},
  \bibnamefont{and} \bibinfo{author}{\bibfnamefont{K.}~\bibnamefont{Wapenaar}},
  \bibinfo{journal}{Geophysical Prospecting} \textbf{\bibinfo{volume}{58 (3)}},
  \bibinfo{pages}{361} (\bibinfo{year}{2010}).

\bibitem[{\citenamefont{Thorbecke and
  Brackenhoff}(2019)}]{jan_thorbecke_2019_3374728}
\bibinfo{author}{\bibfnamefont{J.}~\bibnamefont{Thorbecke}} \bibnamefont{and}
  \bibinfo{author}{\bibfnamefont{J.}~\bibnamefont{Brackenhoff}},
  \emph{\bibinfo{title}{{O}pen{S}ource}} (\bibinfo{year}{2019}),
  \urlprefix\url{https://doi.org/10.5281/zenodo.3374728}.

\end{thebibliography}

\end{document}